\pdfoutput=1

\documentclass[singlespacing, 11pt]{article}

\usepackage{lmodern}
\usepackage{amssymb, amsmath}
\usepackage{ifxetex,ifluatex}
\usepackage[T1]{fontenc}
\usepackage[utf8]{inputenc}

\IfFileExists{microtype.sty}{%
\usepackage{microtype}
\UseMicrotypeSet[protrusion]{basicmath}  % disable protrusion for tt fonts
}{}

\usepackage{color}
\usepackage{xcolor}
\usepackage[margin=1in]{geometry}
\usepackage{pdflscape}

\usepackage{calc}

\usepackage[originalparameters]{ragged2e}  % prevent hyphentation

\renewcommand{\raggedright}{\RaggedRight}

\usepackage{mdwlist}

\usepackage{etoolbox}
\usepackage{setspace}
\AtBeginEnvironment{quote}{\singlespacing}
\AtBeginEnvironment{tabular}{\singlespacing}
\AtBeginEnvironment{longtable}{\singlespacing}
\newlength{\normparskip}
\usepackage{ctable}
\usepackage{booktabs}
\usepackage{longtable}
\usepackage{multirow}

\usepackage{array}

\usepackage{afterpage}

\belowbottomsep=\belowrulesep

\usepackage[export]{adjustbox}  % sizing
\usepackage[section]{placeins}  % provides \FloatBarrier
\usepackage{graphicx}
\graphicspath{{figures/}{../figures/}}

\makeatletter
\def\maxwidth{\ifdim\Gin@nat@width>\linewidth\linewidth\else\Gin@nat@width\fi}
\def\maxheight{\ifdim\Gin@nat@height>\textheight\textheight\else\Gin@nat@height\fi}
\makeatother
\setkeys{Gin}{width=\maxwidth,height=\maxheight,keepaspectratio}

\usepackage[width=\textwidth, justification=centering]{caption}
\usepackage[singlelinecheck=off,justification=centering]{subcaption}
\DeclareGraphicsExtensions{.pdf,.png}

\usepackage{footnote}   % load xcolor first
\makesavenoteenv{tabular}
\makesavenoteenv{table}

\usepackage[flushmargin, bottom]{footmisc}  % http://www.compsoc.man.ac.uk/~moz/www-user/help/footnotes.php

\usepackage[unicode=true]{hyperref}
\hypersetup{breaklinks=true,
            bookmarks=true,
            pdfauthor={Elizabeth Roberto},
            pdftitle={The Spatial Proximity and Connectivity (SPC) Method for Measuring and Analyzing Residential Segregation},
            colorlinks=true,
            citecolor=blue,
            urlcolor=blue,
            linkcolor=black,
            pdfborder={0 0 0}}
\title{{\singlespacing The Spatial Proximity and Connectivity (SPC) Method\\ for Measuring and Analyzing Residential Segregation\footnotemark[1] \\ } \vspace{-2ex}}
\author{Elizabeth Roberto\footnotemark[2]}
\date{August 23, 2017}

\begin{document}

\renewcommand{\thefootnote}{\fnsymbol{footnote}}
\footnotetext[1]{I wish to thank Julia Adams, Richard Breen, Paul DiMaggio, Jacob Faber,
  Angelina Grigoryeva, Jackelyn Hwang, Scott Page, and Andrew Papachristos 
  for their valuable feedback.\\
  This research was supported by the James S. McDonnell Foundation Postdoctoral
  Fellowship Award in Studying Complex Systems, and the The Princeton Institute for
  Computational Science and Engineering (PICSciE) and the Office of Information Technology's
  High Performance Computing Center at Princeton University.}
\footnotetext[2]{Department of Sociology, Princeton University, 107 Wallace Hall, Princeton, NJ 08544}
\renewcommand{\thefootnote}{\arabic{footnote}}

\setlength{\parskip}{0pt plus 0pt}
\maketitle
\setlength{\parskip}{\normparskip}

\phantomsection 
\addcontentsline{toc}{section}{Abstract}
\begin{singlespace}
\begin{abstract}
\normalsize In recent years, there has been increasing attention to the spatial dimensions of residential segregation, such as the spatial arrangement of segregated neighborhoods and the geographic scale or relative size of segregated areas. However, the methods used to measure segregation do not incorporate features of the built environment, such as the road connectivity between locations or the physical barriers that divide groups. This article introduces the Spatial Proximity and Connectivity (SPC) method for measuring and analyzing segregation. The SPC method addresses the limitations of current approaches by taking into account how the physical structure of the built environment affects the proximity and connectivity of locations. In this article, I describe the method and its application for studying segregation and spatial inequality more broadly. I demonstrate one such application---analyzing the impact of physical barriers on residential segregation---with a stylized example and an empirical analysis of racial segregation in Pittsburgh, PA. The SPC method contributes to scholarship on residential segregation by capturing the effect of an important yet understudied mechanism of segregation---the connectivity, or physical barriers, between locations---on the level and spatial pattern of segregation, and enables further consideration of the role of the built environment in segregation processes.
\end{abstract}
\end{singlespace}

\begin{center}
\end{center}

\newpage

\phantomsection 
\addcontentsline{toc}{section}{Introduction}\label{introduction}

Over the last century, scholars have generated a cumulative body of
knowledge on the prevalence, causes, and consequences of residential
segregation, and have established segregation as a key mechanism of
social stratification (for a review, see Charles 2003)(see also Bruch
2014; Logan 2013; Massey 2016; Massey and Denton 1993; Quillian 2012;
Sampson and Sharkey 2008). In recent years, there has been increasing
acknowledgement that popular summary indexes of segregation, like the
dissimilarity index, allow researchers to describe certain
characteristics of residential segregation, but fail to capture
differences in the spatial organization of segregation patterns, such as
the spatial arrangement of segregated neighborhoods and the geographic
scale or relative size of segregated areas (Brown and Chung 2006;
Morrill 1991; Reardon and O'Sullivan 2004; White 1983). Consequently,
there has been more attention to the spatial dimensions of residential
segregation, with newly developed methods and an increasing number of
studies on to the topic (for example, Bischoff 2008; Crowder and South
2008; Farrell 2008; Fischer 2008; Fischer et al. 2004; Folch and Rey
2016; Fowler 2015; Grannis 1998; Grigoryeva and Ruef 2015; Lee et al.
2008; Lichter, Parisi, and Taquino 2015; O'Sullivan and Wong 2007;
Reardon and Bischoff 2011; Reardon et al. 2009; Spielman and Logan
2013).

Although these developments have undoubtedly advanced our understanding
of segregation, the newly developed methods do not incorporate features
of the built environment into the measurement of segregation, such as
the physical barriers that divide cities and reduce connectivity between
nearby areas. Physical barriers influence residential sorting processes
by providing clear divisions between areas and yielding agreement among
residents, real estate agents, and other institutional actors about
where one neighborhood ends and another begins (Ananat 2011; Bader and
Krysan 2015). They ease the categorization of areas in the housing
search process (Bader and Krysan 2015; Krysan and Bader 2009) and
purport to offer ``protection'' from residents on the other side of a
boundary (Atkinson and Flint 2004; Blakely and Snyder 1997; Low 2001;
Schindler 2015). As a result, the presence of these barriers can create
distinct social conditions and experiences for individuals on different
sides of them, as exemplified by the common metaphor, ``the other side
of the tracks.''

Spatial features like streets or landmarks can carry symbolic meaning as
a border between residents of different groups (Kramer 2017) and can
structure residential preferences and discrimination (Besbris et al.
2015). However, they are still fluid and negotiable and offer the
\emph{possibility} for integration (Anderson 1990; Hunter 1974; Hwang
2016; Suttles 1972). In contrast, physical barriers are strong and
persistent forms of boundaries that limit physical connectivity between
areas and they require institutional action to dismantle, such as urban
planning and infrastructure investment (Jackson 1985; Mohl 2002;
Schindler 2015). Moreover, some physical barriers were originally
constructed with an intention to racially segregate nearby populations
(Mohl 2002; Schindler 2015; Sugrue 2005), such as in the selection of
routes for interstate highways built during the 1950s and 1960s in
cities including Chicago and Atlanta (Mohl 2002).

This article introduces a new method for measuring and analyzing
segregation: the Spatial Proximity and Connectivity (SPC) method. The
SPC method is designed to address the limitations of current approaches
by integrating features of the built environment into the measurement of
segregation and analyzing the level and spatial pattern of segregation
revealed by such measures. My proposed approach contributes to the
scholarship on residential segregation by capturing the effect of an
important yet understudied mechanism of segregation---the connectivity,
or physical barriers, between locations---and it enables further
consideration of the role of the built environment in segregation
processes.

In the first section of the article, I review current spatial approaches
for measuring residential segregation and discuss their limitations. I
then introduce the SPC method and describe each of its steps. I outline
a wide range of applications relevant to studying segregation and
spatial inequality more broadly. I demonstrate one such
application---analyzing the impact of physical barriers on residential
segregation---using a stylized example and an empirical case. The first
demonstration uses a stylized city has a north-south pattern of
segregation similar to the patterns of white-black segregation in
St.~Louis or Cleveland. I introduce various physical barriers into the
city and compare how segregation changes when each type of barrier is
present. In the second demonstration, I examine patterns of racial
residential segregation in Pittsburgh, PA. I compare segregation
measures that do and do not take into account the road connectivity
between locations, and I evaluate the extent to which physical barriers
facilitate greater separation between groups and increase the local and
city-wide levels of segregation. In the concluding section, I discuss
possible extensions to the SPC method to address related research
questions.

\section{Measuring Residential
Segregation}\label{measuring-residential-segregation}

Scholars have engaged in a longstanding debate about how best to measure
residential segregation (for a brief history, see Reardon and Firebaugh
2002). The index of dissimilarity came into widespread use in the
mid-20th century and remains the most popular measure of segregation.
Although summary indexes, like the dissimilarity index, allow
researchers to describe distributional characteristics of residential
segregation (Massey and Denton 1988), they are ``aspatial''---they do
not integrate the fundamentally spatial concepts of proximity and
geographic scale into the measurement of segregation. Spatial proximity
concerns where areas are located relative to one another, such as how
neighborhoods are spatially arranged in a city. Geographic scale
concerns the relative size of segregated clusters or the geographic
extent of segregation patterns.

The shortcomings of aspatial approaches are summarized by two well
documented methodological problems: the Checkerboard Problem and the
Modifiable Areal Unit Problem. The Checkerboard Problem (Duncan and
Duncan 1955; Taeuber and Taeuber 1965; White 1983) describes the failure
of aspatial approaches to account for the arrangement or relative
position of spatial units. If we imagine the squares on a checkerboard
to be neighborhoods with a given composition, there are many possible
ways the neighborhoods can be arranged to create different spatial
patterns of segregation. But, if we do not take into account where they
are located relative to one another, any arrangement of the
neighborhoods would result in the same segregation score. The Modifiable
Areal Unit Problem (MAUP) occurs when the composition of spatial units
is affected by changes to the boundaries, number, or size of the spatial
units (Fotheringham and Wong 1991; Openshaw 1984; Openshaw and Taylor
1979). This is problematic if the spatial units are defined by arbitrary
boundaries, rather than being socially meaningful entities (Bischoff
2008; Reardon and O'Sullivan 2004).

Several spatial approaches for measuring segregation have been developed
to address these concerns, which I organize into three types of
strategies: comparing nested levels of geography, identifying spatial
neighbors, and constructing egocentric neighborhoods. In the the
remainder of this section, I briefly describe each of these strategies
and summarize the problems that they solve and those that remain.

First, a popular strategy for addressing the checkerboard problem is to
analyze the geographic scale of segregation patterns by comparing
segregation within or between nested levels of geography. Studies have
measured segregation for increasingly large units of geography, such as
census tracts nested in municipalities within a metropolitan area, and
compared the segregation occurring at each level. For example, Massey,
Rothwell, and Domina (2009) measure segregation with the Dissimilarity
Index and find that declines in black-white segregation have occurred
primarily at the tract-level, with very little change in the segregation
level of cities, counties, or states since 1970. However, using this
approach, we can not compare how much the segregation for each level of
geography contributes to the overall level of segregation in the region.

Another approach is to use a measure that allows one to \emph{decompose}
the overall segregation in a region into the segregation occurring
within and between places in the region, and then compare the
contribution of micro (i.e.~within each place) and macro (i.e.~between
places) segregation components to overall segregation. Several recent
studies have used Theil's entropy index for decompositions within and
between the cities and suburbs of U.S. metropolitan areas (Farrell 2008;
Fischer 2008; Fischer et al. 2004; Lichter et al. 2015; Parisi, Lichter,
and Taquino 2011, Fowler, Lee, and Matthews (2016)). Theil's entropy
index is a desirable measure because it is additively decomposable,
however unlike the Dissimilarity Index, it is a measure of diversity not
segregation. It compares the \emph{diversity} of smaller spatial units
relative to the larger aggregate area, rather than comparing their
\emph{compositions}. Measuring the difference between their
\emph{compositions}, rather than their \emph{diversity}, is preferable
for a segregation index because it allows us to directly assess groups
that are over- or under-represented relative to their overall
proportions, whereas diversity measures necessarily lose information
about the specific group proportions and instead focus on the variety or
relative quantity of groups (Roberto, 2016).

A second strategy for addressing the checkerboard problem is to
integrate information about the relative proximity of neighborhoods
(e.g.~census tracts) when measuring segregation. This is typically done
in one of two ways: either by identifying adjacent neighborhoods (i.e.,
immediate neighbors that share a boundary, neighborhoods that have a
common neighbor, etc.), or by calculating the geographic distance
between the center point of each pair of neighborhoods. Several spatial
indexes have been developed to accommodate information about proximity
when measuring segregation (for a review, see Reardon and O'Sullivan
2004). They typically use a proximity function to incorporate the
population of nearby areas into each neighborhood's population
composition, with a weight that determines the relative contribution of
distant versus nearby areas. For example, a uniform (or rectangular)
proximity function gives equal weight to distant and nearby areas, as
long as they are within a given distance band (e.g., Jargowsky and Kim
2005; Wu and Sui 2001), or a distance-decay function can be used to
weight nearby areas more heavily than distant areas, with the rate of
decay intended to represent the influence of distance on social
interaction patterns (White 1983). Although this strategy accounts for
the spatial arrangement of neighborhoods, it still relies on census
units, such as tracts, which vary in geographic size and population
density across the country.

A third strategy proposed by Reardon and colleagues (Lee et al. 2008;
Reardon et al. 2009, 2008) uses ``egocentric neighborhoods'' to measure
segregation and addresses both the checkerboard problem and MAUP. In
contrast to conventional approaches that use census tracts to measure
segregation, they superimpose a grid with 50 by 50 meter cells over the
census blocks of a metropolitan area and estimate the population in each
cell.\footnote{They estimate the population count and composition of
  each cell using Tobler's (1979) pycnophylactic (``mass preserving'')
  smoothing method, which reduces differences in the population of
  neighboring cells that fall along the boundary of a block.} They then
measure the shortest straight line distance between the center points of
all pairs of cells, and use these distances to construct local
environments, or ``egocentric neighborhoods,'' around each of the cells.
The local environment of each cell includes nearby cells within a
particular distance, and they systematically vary the distance using
radii of .5, 1, 2, and 4 km (.3, .6, 1.2, and 2.5 miles). They use a
proximity function that weights the share of the population of nearby
cells that will be included in a cell's local environment.\footnote{They
  measure proximity with a distance-decay function: a two-dimensional
  biweight kernel proximity function. Their proximity function is
  bounded by the radius of local environments, and areas beyond that
  distance receive a weight of 0. Similar to a Gaussian kernel, but
  bounded for each radius to reduce computation (Reardon et al. 2008).
  They exclude areas outside the metropolitan area, even if they are
  within a cell's local environment.} They measure segregation
separately for local environments constructed with each radius and
compare changes in the level of segregation as the radius increases.

By dispensing with census tracts in favor of egocentric neighborhoods of
various sizes, this approach is able to distinguish between the
\emph{geographic} and \emph{methodological} scale of
segregation---between the scale at which segregation is experienced in
social environments and the level of aggregation in the data (Reardon et
al. 2008). Although this is a large step forward for spatial segregation
measurement, a notable limitation remains: the method does not integrate
features of the built environment into the measurement of distance.
Specifically, by using the straight line distance between grid cells to
measure proximity, the approach ignores the physical barriers that
divide urban space and the connectivity provided by roads. It is
therefore unable to detect any difference in segregation whether nearby
areas are separated by a physical barrier, such as a fence, railroad
tracks, or dead-end streets, or well connected by roads.

\section{The Spatial Proximity and Connectivity (SPC) Method for
Measuring and Analyzing
Segregation}\label{the-spatial-proximity-and-connectivity-spc-method-for-measuring-and-analyzing-segregation}

I propose a new method for measuring and analyzing segregation, the
Spatial Proximity and Connectivity (SPC) method. Consistent with recent
advancements in segregation measurement, SPC addresses the Checkerboard
Problem and MAUP by measuring spatial proximity and comparing
segregation at multiple geographic scales. However, SPC also addresses
the limitations of previous approaches by using a realistic measure of
distance that integrates information about the built environment.

Existing methods that use distance to measure spatial proximity rely on
straight line distance---the shortest distance from Point A to Point
B---without considering that spatial areas are often not connected in
this way but rather through a road network. In contrast, SPC measures
the shortest distance between all residential locations along a city's
road network, which reflects the connectivity between locations and the
separation imposed by physical barriers. This is an important feature of
SPC because two residential areas may be spatially proximate to each
other, but not well connected by roads (Neal 2012). For example, In a
study of racial settlement patterns in Los Angeles and San Francisco,
Grannis (1998) found that connectivity along small, residential streets
was more important than mere proximity in predicting racial segregation
patterns.

Physical barriers have also been used as mechanisms to reinforce or
exacerbate segregation by facilitating greater separation between
ethnoracial groups in nearby areas. For example, in ``Crabgrass
Frontier'', Jackson (1985) describes adjacent black and white
neighborhoods in the vicinity of Eight Mile Road in Detroit in the late
1930s. None of the white families could get Federal Housing
Administration (FHA) mortgages ``because of the proximity of an
`inharmonious' racial group'' (p.~209). After a developer built a
concrete wall between the neighborhoods in 1941, FHA approved mortgages
for properties in the white neighborhood. The SPC method contributes to
the scholarship on residential segregation by capturing the effect of
this additional mechanism of segregation---the connectivity, or physical
barriers, between locations---on the level and pattern of segregation.

In this section, I describe each step of the SPC method. Using road
distance to measure the proximity and connectivity between locations
requires six steps: 1) linking the geographic data for blocks and roads
2) estimating the population count and composition at locations on the
road network, and 3) calculating the distance of the shortest path
between all locations, 4) constructing local environments around each
location, 5) calculating proximity weights, and 6) measuring
segregation.

For each of these steps, I use R software (R Core Team 2014) and add-on
packages designed for working with spatial data (Bivand and Rundel 2014;
Bivand, Keitt, and Rowlingson 2014; Neuwirth 2014; Pebesma and Bivand
2015), networks (Csardi and Nepusz 2006), and large matrices (Kane,
Emerson, and Haverty 2013; Revolution Analytics and Weston 2014).
Although my explanation focuses on using the SPC approach to study
residential segregation in cities, the method is applicable to any area
of interest (e.g.~school districts, metropolitan areas, states, etc.),
including rural areas.

\subsection{1. Link the Geographic Data}\label{link-the-geographic-data}

SPC uses publicly available population data from the 2010 decennial
census (U.S. Census Bureau 2011) and the geographic data provided in
TIGER/Line shapefiles (U.S. Census Bureau 2012). The U.S. census
subdivides the entire U.S. into several nested geographic units. I use
population data for census blocks---the smallest unit of census
geography. Blocks are polygons that are typically bounded by street or
road segments on each side and typically correspond to a residential
city block in urban areas. Blocks can also represent spatial areas
without population or non-residential land use, such as industrial
areas, parks, or areas between railroad tracks. Blocks are nested within
census tracts---the most commonly used unit of census geography for
measuring segregation, which contain an average population of 4,000
individuals.

SPC uses the TIGER/Line shapefiles for ``faces'' and ``edges'' to define
the geographic boundaries of blocks and the path of roads. Faces are
polygons that represent area features, such as blocks. Each face is
assigned a permanent unique identifier (uid) by the Census Bureau. In
most cases a block consists of a single face, but in some cases a block
may contain two or more faces (e.g., if an alley subdivides a block).
Each face is bounded by one or more edges. Edges are lines features,
including road segments, and each edge has a uid. Each edge is
associated with two faces---one on each of its sides. The two end points
of an edge are called nodes, and each has a uid. A single node may be
associated with multiple edges, such as a node that joins two road
segments together.

\begin{figure}[h]
  \includegraphics[width=5.25in, center]{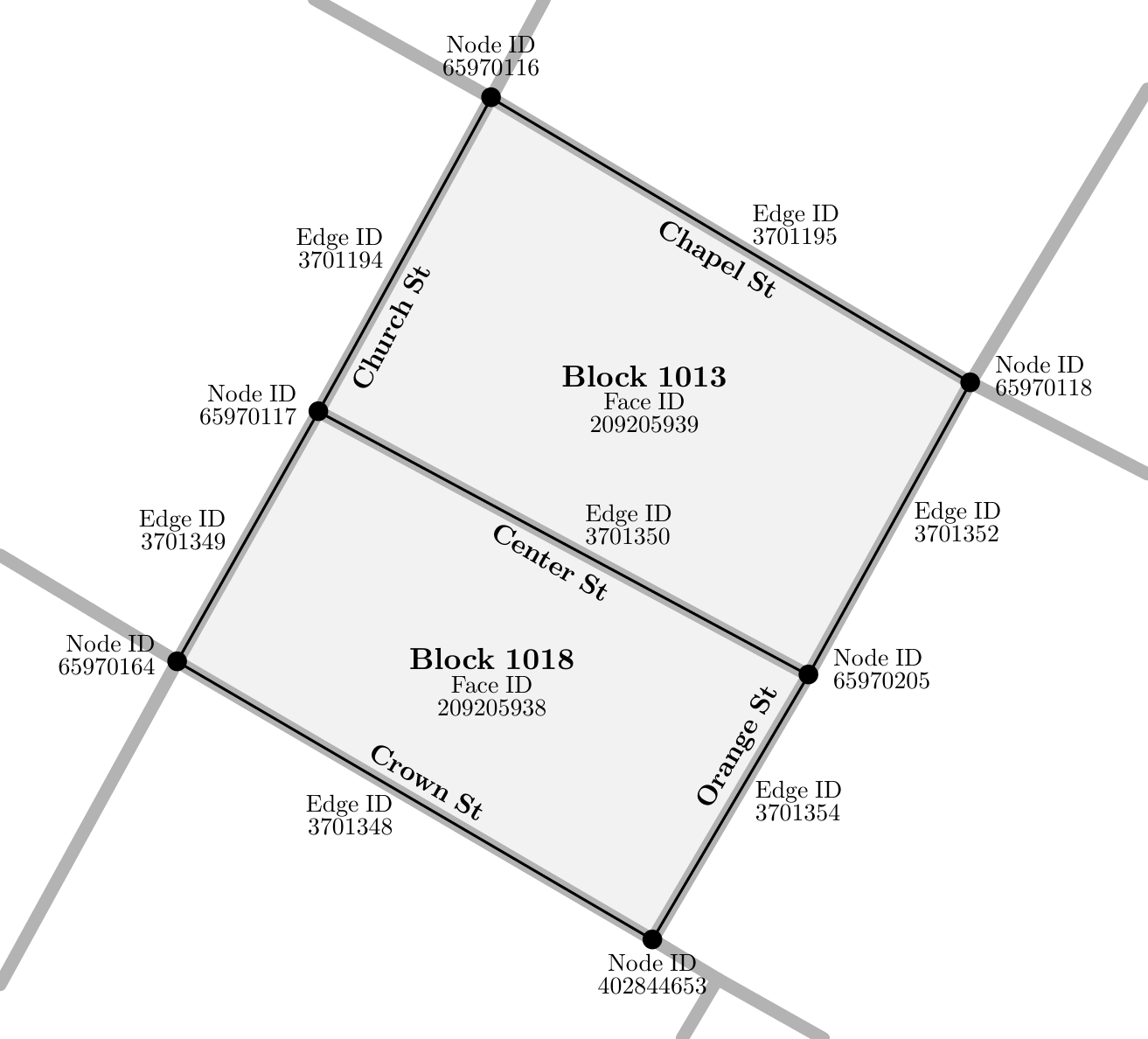}
  \caption{Example of the TIGER/Line Features used to Construct Road Networks}
  \label{fig:topo1}
\end{figure}

For example, Figure \ref{fig:topo1} shows two blocks, the seven road
segments that define their perimeters, and the intersections of the
roads. Each of the blocks has one face, each with a uid. Each road
segment has an edge uid, and each end point of a road has a node uid.
Roads that intersect have nodes in common. For example, node 65970117 in
Figure \ref{fig:topo1} is an end point of edge 3701349, 3701194, and
3701194. This node is the shared intersection of Center St.~and the two
segments of Church St.

\begin{center}
[Figure \ref{fig:topo1} about here.]
\end{center}

I use the uids to link the geographic data for each city by identifying
the relationships between blocks, faces, edges, and nodes. The data
record for each face includes its uid, and if it represents a block
feature, then it includes the uid for the block. The data record for
each edge includes: the edge's uid, the two node uids for its end
points, and the two face uids for its sides. The record also indicates
whether the edge is a road feature, and if so it provides a
classification code for the type of road feature (e.g., primary road,
local road, alley, etc.). For each block, I identify the face uid(s)
associated with the block, find the uid for any roads features
(including alleys and pedestrian walkways) that have the block's face
uid(s) listed as one of its sides, and collect all of the node uids
associated with those roads. The result is a list of the road uids and
node uids associated with each block.

\subsection{2. Estimate the Population Count and Composition of
Nodes}\label{estimate-the-population-count-and-composition-of-nodes}

Once the geographic data for blocks, roads, and nodes have been linked,
I estimate the population count and composition at each of the nodes.
This procedure distributes the aggregate population of each block to
point locations on roads by assigning a portion of each block's
population to the nodes associated with the block.\footnote{If one had
  access to addresses-level data, this estimate procedure would not be
  necessary. Instead, SPC could use the population count and composition
  at each address or, to reduce computational demands, the address-level
  populations could be assigned to the nearest node.} Later, this will
allow us to calculate the population composition in the local
environment around each node. It also has the advantage of removing the
arbitrary administrative boundaries of individual census blocks, and it
smoothes the distribution of the population and sharp discontinuities
that may occur along the administrative boundaries.

I assign the block's population to the nodes in two steps. First, I
assign individuals to one of the roads associated with the block, with
the probability of assignment equal to the length of the road segment.
Second, I randomly assign individuals to one of the two nodes that are
the end points of their assigned road segment. When adjacent blocks are
associated with the same node, such as node 65970117 in Figure
\ref{fig:topo1}, the node will likely receive a portion of each block's
population.

The random assignment of block populations to nodes will affect the
population count and composition of each node. The randomness of the
procedure would likely affect segregation levels if I were measuring
segregation aspatially and using each node as a unit of analysis.
However, SPC measures segregation in the local environments around each
node, as I will describe in the fourth step, which incorporates nearby
populations into the composition. Even at a reach of 0 km, much of the
variability of random assignment is mitigated, because adjacent blocks
share nodes and each block contributes to the intersection's population.
To err on the side of caution, I suggest that the size of local
environments should be at least as large as the typical census block in
a given city, in which case variability in the population count or
composition due to sampling is likely to be minimal.

\subsection{3. Calculate the Shortest
Paths}\label{calculate-the-shortest-paths}

The next step of the SPC method is to measure the shortest path between
all pairs of nodes. The length of the shortest paths is the minimum road
distance between nodes. I measure the shortest paths by constructing a
graph that represents the road network.

SPC takes advantage of the relational nature of the geographic data to
construct the graph. The edgelist of the graph contains each road
segment as an edge and its end points---the nodes---as the vertices. A
single node can join multiple road segments, which provides the
necessary linkages to construct the network. The record for each edge
includes: the uid for the road segment (i.e., the edge), the uids for
the nodes at its end points (i.e., the vertices), and the length of the
road segment (i.e., the edge weight). The graph is undirected, meaning
that the if vertex A is connected to vertex B, then vertex B is also
connected to vertex A.\footnote{A directed graph can be used with the
  SPC method if, for example, one wanted to represent the differences in
  connectivity provided by one-way vs.~two-way streets.} The weight of
the edge connecting vertex A and B represents the road distance between
them. Once the graph is constructed, I calculate the length of the
shortest path between each pair of nodes in the network using the
Dijkstra algorithm implemented in the igraph package for R (Csardi and
Nepusz 2006).

\subsection{4. Construct Local
Environments}\label{construct-local-environments}

I use the ``egocentric neighborhoods'' strategy described earlier to
construct local environments around each node, systematically vary their
size, and record the population composition within all local
environments of each size (Lee et al. 2008; Reardon et al. 2009, 2008).
However, I define the size, or \emph{reach} (i.e., the distance in each
direction from a given location), of local environments using the road
network distance, rather than straight line distance. I systematically
vary the reach of local environments within a range of values, such as
.1 to 10 km (.062 to 6.2 miles). Local environments with a reach of .1
km are about the size of a block in many cities, whereas local
environments with a reach of 10 km encompass a substantial portion of
all but the largest U.S. cities.

Local environments can span bodies of water, such as a rivers and lakes,
and will include the population on the other side of the water if it is
within the reach. For locations near the boundary of a city, the local
environments are constrained to be within the boundary. For example, if
the reach is 1 km, the local environment of an node within 1 km of the
city boundary does not include residents across the city's boundary,
even if they are within 1 km of the node. If a city is bordered by
another country (e.g., where Detroit borders Canada), the local
environments do not cross the U.S. border.

\subsection{5. Calculate Proximity
Weights}\label{calculate-proximity-weights}

I use a uniform proximity function to weight the relative contribution
of distant vs.~nearby nodes in the population of each node's local
environment (e.g., Jargowsky and Kim 2005; Wu and Sui 2001). For each
node, the uniform (or rectangular) proximity function gives a weight of
one to other nodes that are within the reach of the local environment
and a weight of zero to any nodes that are outside the reach of the
local environment. A uniform proximity function is calculated as:
\[ \phi(i,j)=\begin{cases} 1 & \text{if } d(i,j)<r \\ 0 & \text{otherwise} \end{cases}\]
where \(d(i,j)\) is the pairwise distance between nodes \(i\) and \(j\)
and \(r\) is the reach of local environments.

Alternatively, a biweight proximity function can be used to weight the
relative contribution of nodes within the local environment, with nearby
areas weighted more heavily than distant areas (White 1983). A biweight
proximity function is calculated as:
\[ \phi(i,j)=\begin{cases}\begin{bmatrix} 1-\left(\frac{d(i,j)}{r}\right)^{2} \end{bmatrix}^{2} & \text{if } d(i,j)<r \\ 0 & \text{otherwise} \end{cases}\]
where \(d(i,j)\) is the pairwise distance between nodes \(i\) and \(j\)
and \(r\) is the reach of local environments.

\subsection{6. Measure Segregation}\label{measure-segregation}

I record the population composition within all local environments, and I
use this information to measure the segregation in each local
environment and the city as a whole. I measure segregation with the
Divergence Index (Roberto, 2016), which measures the difference between
the population composition of each local environment and the city's
overall composition. The Divergence Index measures the same concept of
segregation as the Dissimilarity Index, but it has the several
advantages. The index can be calculated for both discrete and continuous
data, as well as for joint distributions (e.g., income by race). It is
also additively decomposable, which allows one to compare how much of
the overall segregation in a city occurs within vs.~between areas within
the city.

The values of the Divergence Index represents how surprising the
composition of a local environment is, given the overall population
composition of the city. The divergence index equals 0---its minimum
value---when there is no difference between the local and overall
population composition, whereas greater differences produce higher
values and indicate a greater degree of segregation. Local values of the
divergence index will reach their maximum value when the smallest group
in a city is 100 percent of the local population.

The divergence index for location (i.e., node) \(i\)'s local environment
with a reach of \(r\) km is:
\[\tilde{D}_{ri}=\sum_{m}{\tilde{\pi}_{rim}\log{\cfrac{\tilde{\pi}_{rim}}{\pi_m}}}\]
where \(\pi_m\) is group \(m\)'s proportion of the region's (e.g., a
city's) overall population, and \(\tilde{\pi}_{rim}\) is group \(m\)'s
proportion of the proximity weighted population in location \(i\)'s
local environment with reach \(r\). The proximity weighted population
composition for each local environment is calculated as:
\[\tilde{\pi}_{rim} = \frac{\int_{j \in K} \tau_{jm} \phi(i,j)dj}{\int_{j \in K} \tau_{j} \phi(i,j)dj}\]
where \(\tau_{j}\) and \(\tau_{jm}\) are the total and group-specific
population counts for each location \(j\) in region \(K\), and
\(\phi(i,j)\) is the proximity function for locations \(i\) and \(j\).

A region's overall segregation for a given reach of local environments
is the population weighted mean of the divergence index for all
locations, calculated as:
\[\tilde{D}_{r}=\sum_{i}{\frac{\tau_{i}}{T}\tilde{D}_{ri}}\] where \(T\)
is the region's overall population count, and \(\tau_i\) is the
population count in location \({i}\). If all local environments have the
same composition as the region's overall population, then
\(\tilde{D}_{r}\) equals 0, indicating no segregation. More divergence
between overall and local proportions indicates more segregation.

The divergence index can also be used to calculate group-specific
segregation results. For each reach of local environments, the average
degree of segregation experienced by each group is calculated as:
\[\tilde{D}_{rm}=\sum_{i}{\frac{\tau_{im}}{\tau_{m}}\tilde{D}_{ri}}\]
where \(\tau_{m}\) is the region's overall population of group \(m\),
and \(\tau_{im}\) is the population of group \(m\) in location \({i}\).
The weighted mean of the group-specific segregation results is equal to
the region's overall segregation:
\[\tilde{D}_{r}=\sum_{m}{\frac{\tau_{m}}{T}\tilde{D}_{rm}}\]

\section{Applications of the SPC
Method}\label{applications-of-the-spc-method}

The SPC method can be applied in a variety of ways to study segregation
or spatial inequality more broadly. It can measure residential
segregation in cities or any other municipal divisions of interest, such
as metropolitan areas or school districts. Or it can be used to compare
segregation in the vicinity of public institutions (e.g., libraries) and
recreational spaces (e.g., parks) and evaluate the potential for these
places and spaces to bring together a representative mix of the city's
population. It can also measure segregation in the vicinity of
environmental hazards (e.g., hazardous waste sites) to evaluate the
extent to which particular groups are disproportionately exposed to
these risks. Additional data about crime, health, or other population or
environmental characteristics can also be incorporated into the SPC
method for measurement or analysis of multiple spatial attributes.

SPC can be used as a replacement for current methods of measuring
segregation, or it can be used along side straight line distance
segregation measures to evaluate the difference in segregation due to
connectivity and physical barriers. In the remainder of the paper, I
will describe this application of the method and provide two
demonstrations. First, I compare how segregation differs in a stylized
city when various types of physical barriers spatially separate two
groups. Next, I will provide an empirical application that examines the
local and city-wide levels of racial segregation in Pittsburgh, PA.

\subsection{Measuring the Impact of Physical Barriers on Residential
Segregation}\label{measuring-the-impact-of-physical-barriers-on-residential-segregation}

Physical barriers are material structures---such as rivers, parks, or
highways---that reduce the connectivity between locations on either side
of the barrier. Features of a city's street design can act as physical
barriers: dead end streets and cul-de-sacs create excess distance
between locations, whereas a regular street grid provides greater
connectivity between locations. The Dan Ryan Expressway in Chicago's
South Side is a classic example of a physical barrier. It was
constructed in the 1960's and separated the white and black residents on
either side of the highway. Using straight line distance to measure the
proximity of these communities would represent their nearness, but not
their disconnection. Their distance apart would be measured as the width
of the highway---the same as if the highway was never constructed. Using
road network distance more accurately represents the highway as a source
of separation: it is a physical barrier that divides the communities and
facilitated racial segregation, not residential integration.

The difference between the road distance and the straight line distance
between two locations reveals the extent to which road connectivity is
limited or physical barriers are present between locations. The road
distance between any two locations is always equal to or greater than
the straight line distance between them.~ In a city with a regular
street grid with diagonal avenues at every intersection, there would be
no difference between the two distance measurements. ~Even without
diagonal avenues, there would still be very little difference between
the two distance measurements, especially for relatively nearby
locations. If the road network is less connected or if there are other
types of physical barriers present, the road distance between locations
will be greater than their straight line distance. ~ For example, the
presence of a dead-end or cul-de-sac can affect the road distance
between nodes in the area, but it will have no affect on the straight
line distance between the nodes.

To evaluate the impact of connectivity and physical barriers on local
and overall levels of segregation, I compare segregation measures that
use straight line distance and road network distance. Following the
steps of the SPC method, I construct local environments for every node
in a city. However, I define the reach of local environments in two
ways: using straight line distance and using road distance. In both
cases, I systematically vary the reach of local environments from .1 to
10 km (6.2 miles). The presence of physical barriers will only affect
the areas included in the local environments constructed with road
distance, not those constructed with straight line distance. Therefore,
a greater prevalence of physical barriers will result in a greater
difference between the areas included in local environments constructed
with road distance and the areas included in local environments
constructed with straight line distance, and potentially a difference in
their population compositions.

\begin{figure}

  \begin{subfigure}{\textwidth}
    \caption{Local Environment Constructed with Straight Line Distance}
    \label{fig:rle_pt2}
    \includegraphics[height=3.5in, center]{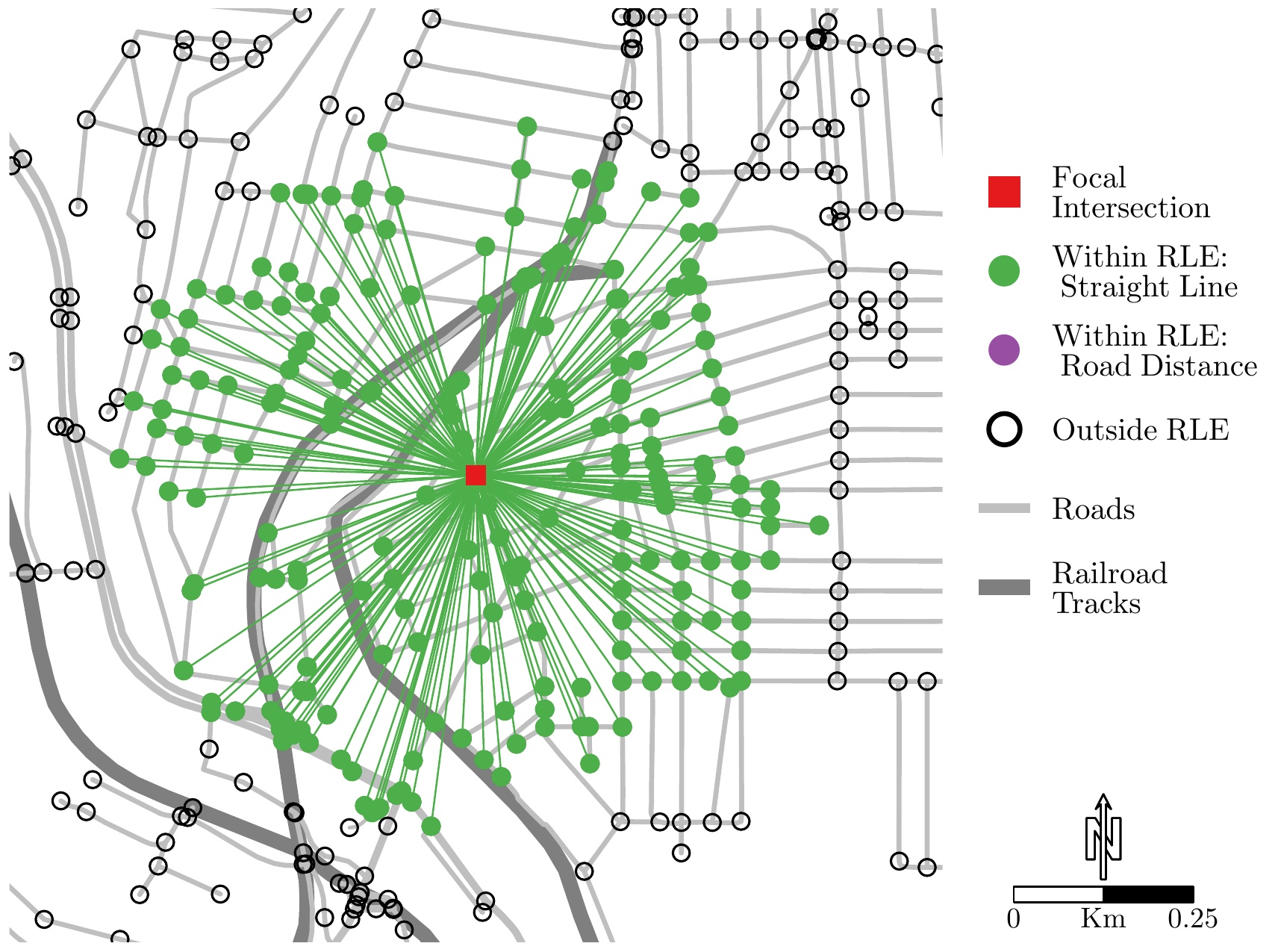}
  \end{subfigure}
  ~\\
  ~\\
  \begin{subfigure}{\textwidth}
    \caption{Local Environment Constructed with Road Network Distance}
    \label{fig:rle_pt3}
    \includegraphics[height=3.5in, center]{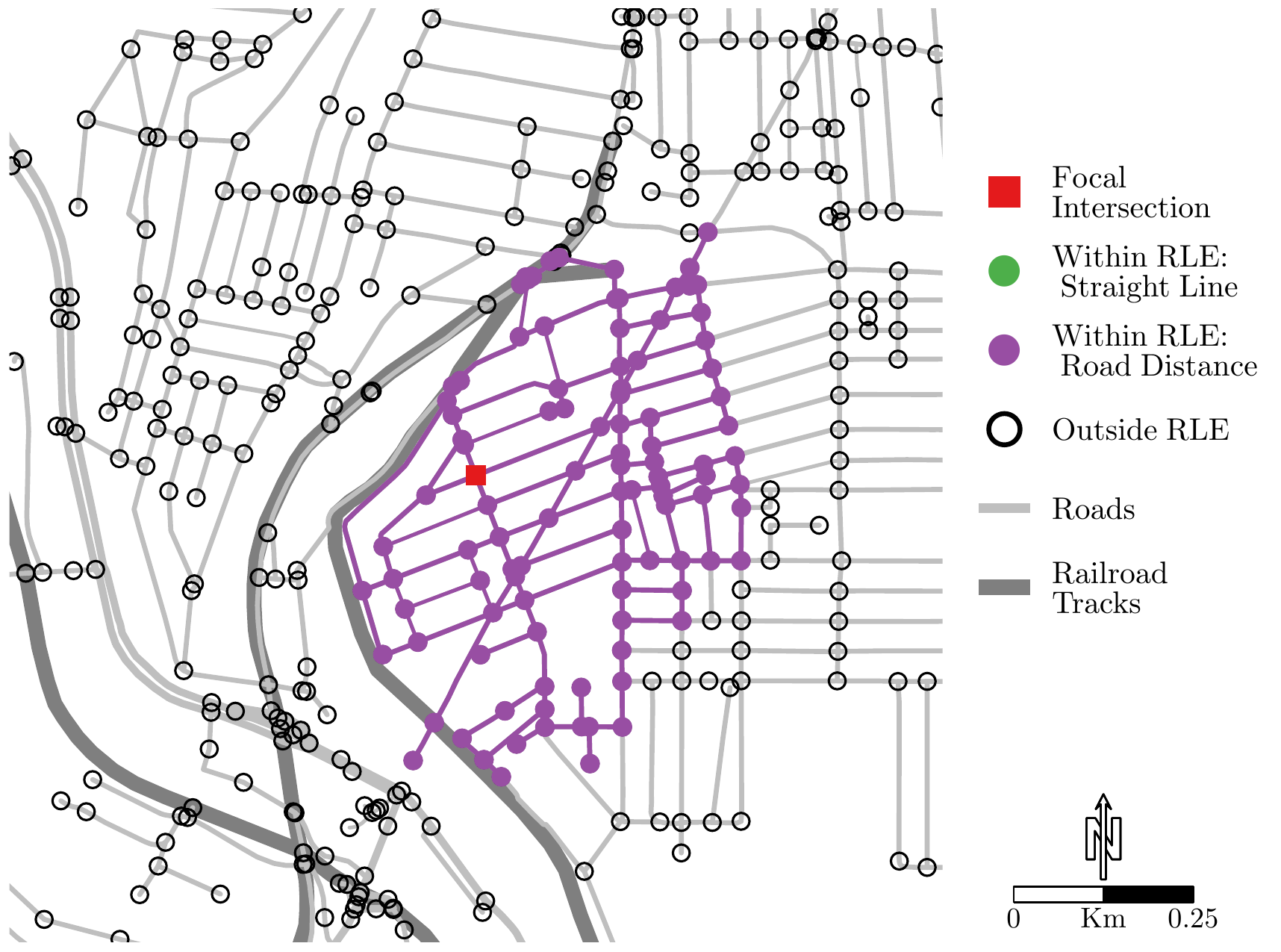}
  \end{subfigure}
  
  \caption{Comparing a Local Environment Constructed with\\ Straight Line Distance and Road Distance\\ (Reach of the Local Environment = .5 km)}
  \label{fig:rle_pt}
\end{figure}

If roads connect all nodes, then the local environments constructed with
each distance measure would be identical in size. The biggest
differences will occur in areas where one or more nodes are not well
connected to other nearby nodes. For example, if railroad tracks create
a physical barrier between nearby areas, they would affect the areas
includes in each node's local environment when they are constructed with
road distance. Figure \ref{fig:rle_pt} illustrates the difference
between a local environment constructed with straight line distance and
a local environment constructed with road distance for one node located
near railroad tracks. Figure \ref{fig:rle_pt2} shows the node's local
environment constructed with straight line distance: all intersections
within .5 km are included in the node's local environment. Figure
\ref{fig:rle_pt3} shows the node's local environment constructed with
road distance. The railroad tracts limit the connectivity to areas west
of the tracks and severely reduce the number of nodes that are included
in the local environment.

\begin{center}
[Figure \ref{fig:rle_pt} about here.]
\end{center}

The presence of physical barriers alone does not indicate that they have
an impact on the level of segregation. For physical barriers to impact
segregation levels, they must create greater separation between areas
with different population compositions. To make this assessment, I
record the population composition within all local environments
constructed with each type of distance for each reach and I measure the
segregation of all local environments using the Divergence Index
(Roberto, 2016), as described earlier. For each reach, I measure
segregation separately for the local environments constructed with each
type of distance.

If there are no physical barriers between locations, the local
environments constructed with straight line distance and those
constructed with road distance will encompass the same areas and have
the same composition. There will be no difference in the level of
segregation for local environments of a given reach constructed with
each type of distance.

If physical barriers are present in an area but there is no difference
between the road distance and straight line distance segregation
measures for the nodes in that area, this indicates that the racial
compositions of their local environments measured by each type of
distance are identical and physical barriers do not structure the
spatial pattern of segregation in that area. If the road distance
segregation measure for a local environment is lower than the straight
line distance segregation measure, this indicates that the local
environment is in an area with a composition that is similar to the city
(i.e., lower segregation) and a physical barrier separates it from an
area with a composition that differs from the city (i.e., higher
segregation).

If the road distance segregation measure for a local environment is
higher than the straight line distance segregation measure, this
indicates that physical barriers play a role in spatially structuring
segregation patterns. The greater the difference, the greater the extent
to which physical barriers divide groups and facilitate segregation.
These differences may be greater for certain reaches of local
environments than others, which would indicate that physical barriers
play a larger role in structuring segregation patterns at some
geographic scales than at others.

\subsubsection{Physical Barriers and Racial Segregation in A
Stylized
City}\label{physical-barriers-and-racial-segregation-in-a-stylized-city}

In this section, I use a stylized city to demonstrate how the SPC method
can be used to evaluate the impact of physical barriers on segregation
levels. The stylized city is much simpler than a real city, in terms of
both its geography and the distribution of the population. I use this
simple example to illustrate the steps of the method and provide a
sample of the results it produces.

The stylized city is calibrated to approximate the population and census
geography of a medium-sized U.S. city or several neighborhoods within a
larger city. It has a population of 500,000 people, and for simplicity,
the population includes only two race groups---white and black. The city
contains 100 tracts, each with a population of 5,000 people. Each tract
contains 25 blocks, with an equal population count of 200 people. Blocks
are bounded by streets, and the length of each side of a block and of
each of the streets segments surrounding it is 250 meters (.16
miles).\footnote{In U.S. cities, blocks vary in size. A standard block
  in Manhattan is about 80 meters by 270 meters. To compare the
  dimensions of a typical block across cities, see:
  http://www.thegreatamericangrid.com/dimensions.} All adjacent streets
are connected, unless a physical barrier is present.

\begin{figure}
  \includegraphics[height=3.5in, center]{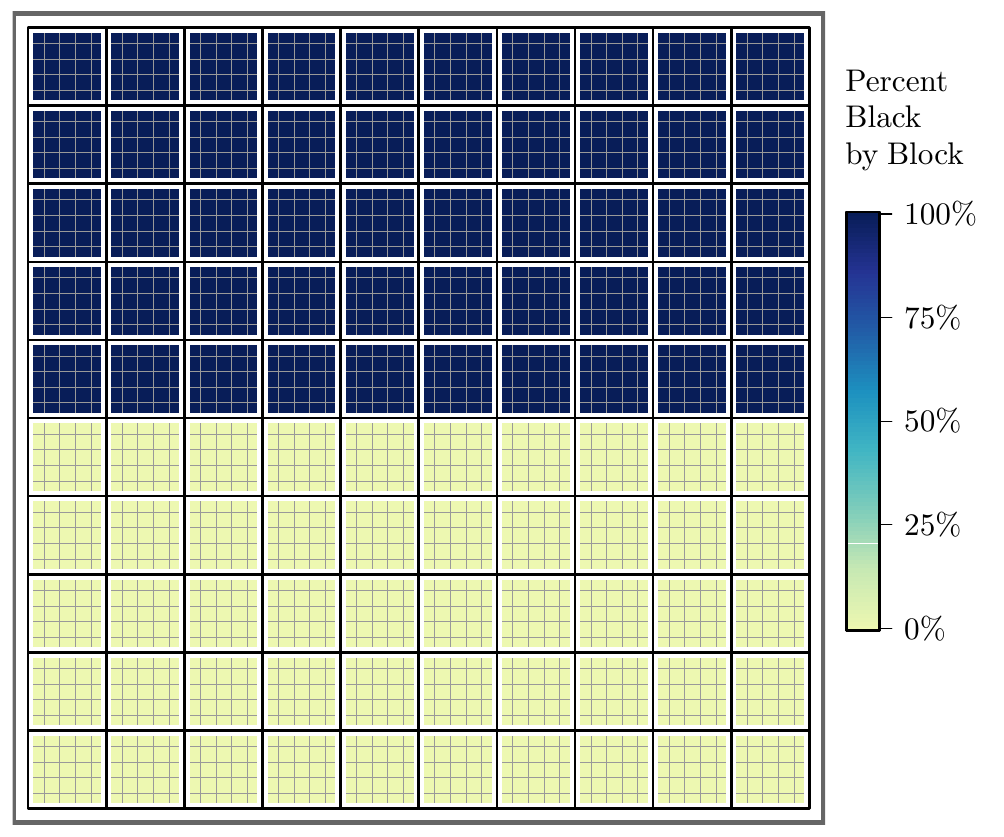}
  \caption{Map of a Stylized City\\ Percent Black in Blocks}
  \label{fig:half_div0_block}
\end{figure}

The city has a spatial pattern of segregation similar to the north-south
patterns of white-black segregation in cities such as St.~Louis and
Cleveland. Figure \ref{fig:half_div0_block} shows a map of the racial
composition of blocks in the stylized city. The thin black lines
indicate the borders of blocks, and the thick black lines indicate the
borders of tracts. The population of the city includes residents who are
white and black, and the fill color of the blocks indicates the percent
black in each block, with darker colors indicating a higher percent. The
population of each block and tract is exclusively either white or black.

\begin{center}
[Figure \ref{fig:half_div0_block} about here.]
\end{center}

I separately introduce three types of physical barriers into the city
and evaluate the impact of each barrier on the level of segregation. The
types of barriers I use are: a barrier that fully divides the north and
south sides of the city, a barrier that spans half the width of the
city, and a segmented barrier that resembles a river with bridges every
few kilometers. Although it is unusual to observe a barrier completely
dividing a city in half, such barriers can be found between
neighborhoods or larger areas within cities.

I use the SPC approach to measure segregation in the city, using both
straight line distance and road network distance to measure the reach of
local environments. I then analyze the differences between the two sets
of results to evaluate the impact of each type of barrier on the level
of segregation.

\begin{figure}
  \includegraphics[height=3.5in, center]{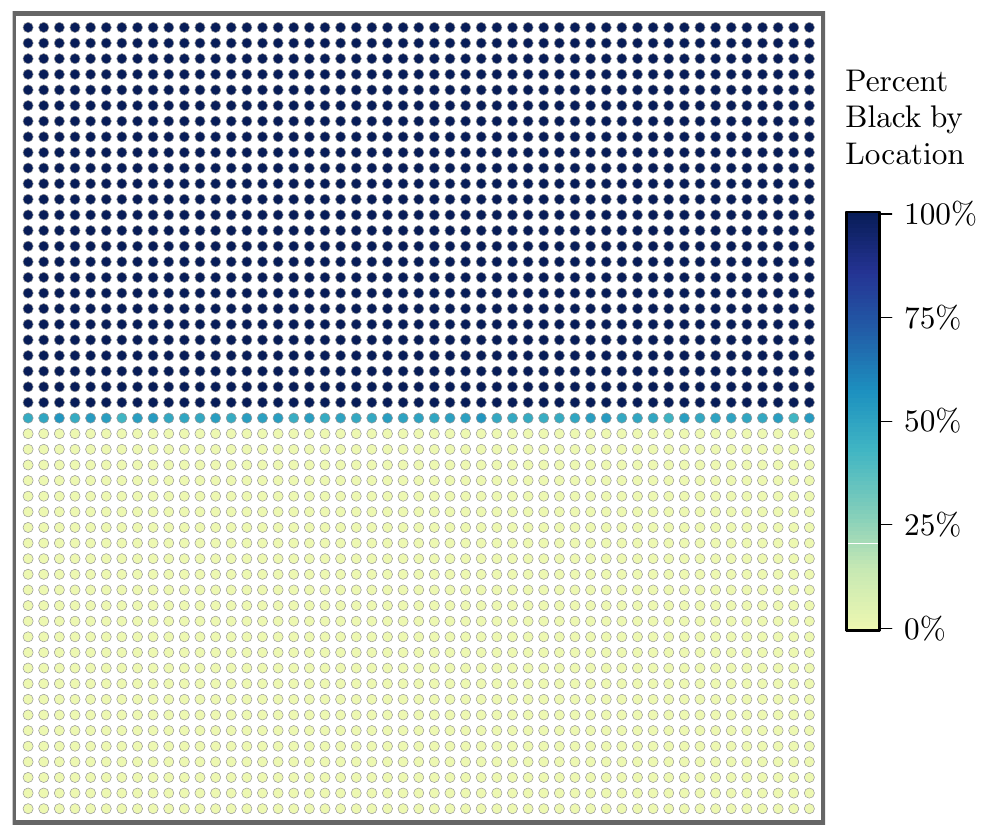}
  \caption{Map of a Stylized City\\ Percent Black in Node Locations}
  \label{fig:half_div0_pts}
\end{figure}

Following the first two steps of the SPC method, I link the geographic
data for blocks and roads and estimate the population count and
composition at each of the nodes (i.e., the intersections or roads).
Figure \ref{fig:half_div0_pts} shows the result of these two steps---the
racial composition of each node. Although each half of the city has a
monoracial population, locations along the midline where the clusters
meet are diverse. The adjacent blocks share roads and intersections, and
individuals living in different blocks on either side of a road are
assigned to one or more of the same nodes.

\begin{center}
[Figure \ref{fig:half_div0_pts} about here.]
\end{center}

Third, I calculate the shortest path along the road network between all
pairs of nodes in the city. I also calculate the straight line distance
between all pairs of nodes. I then construct local environments around
each node using both the road distance and straight line distance
measures. I vary the reach of the local environments, ranging from .1 to
10 km. I calculate the proximity weighted population composition in the
local environment of each node, separately for each distance measure and
each reach. Finally, I use the Divergence Index to measure segregation
in the local environment of each node, calculating separate results for
each distance measure and each reach.

\begin{figure}

  \begin{subfigure}{\textwidth}
    \caption{Road Distance and Straight Line Distance Segregation Measures by Barrier Type}
    \label{fig:ResultsD_half}
    \includegraphics[width=3.9in, center]{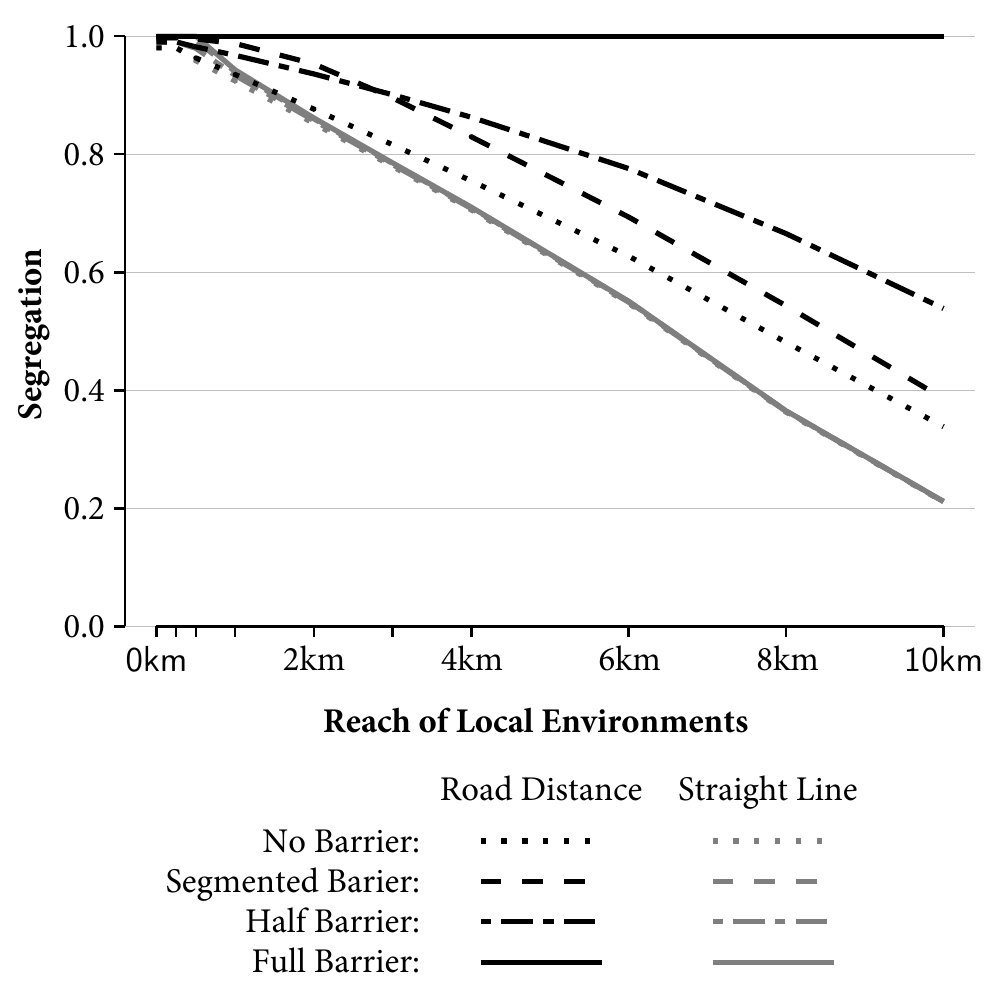}
  \end{subfigure}
  ~\\
  ~\\
  \begin{subfigure}{\textwidth}
    \caption{Difference between Road Distance and Straight Line Distance Segregation Measures by Barrier Type}
    \label{fig:CompareD_half}
    \includegraphics[width=3.9in, center]{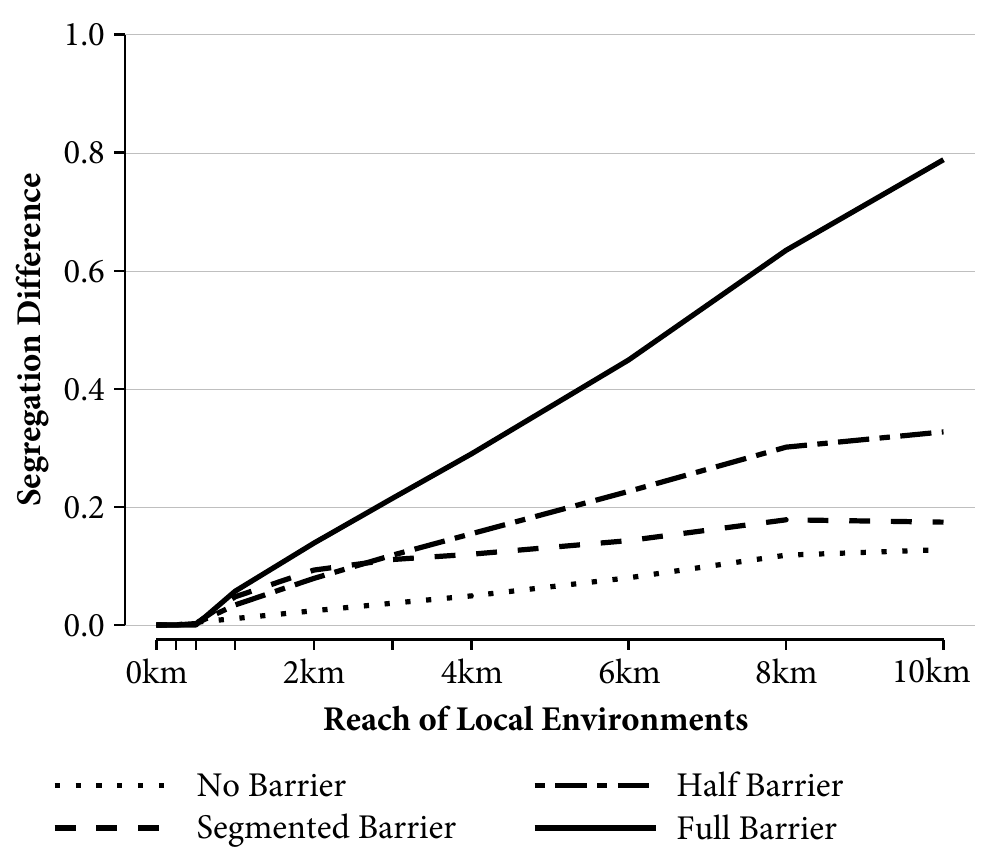}
  \end{subfigure}

  \caption{Segregation in a Stylized City}
  \label{fig:SegD_half}
\end{figure}

\begin{figure}

  \begin{subfigure}[t]{0.49\textwidth}
    \caption{No Barrier}
    \label{fig:half_div0}
    \includegraphics[width=\linewidth, center]{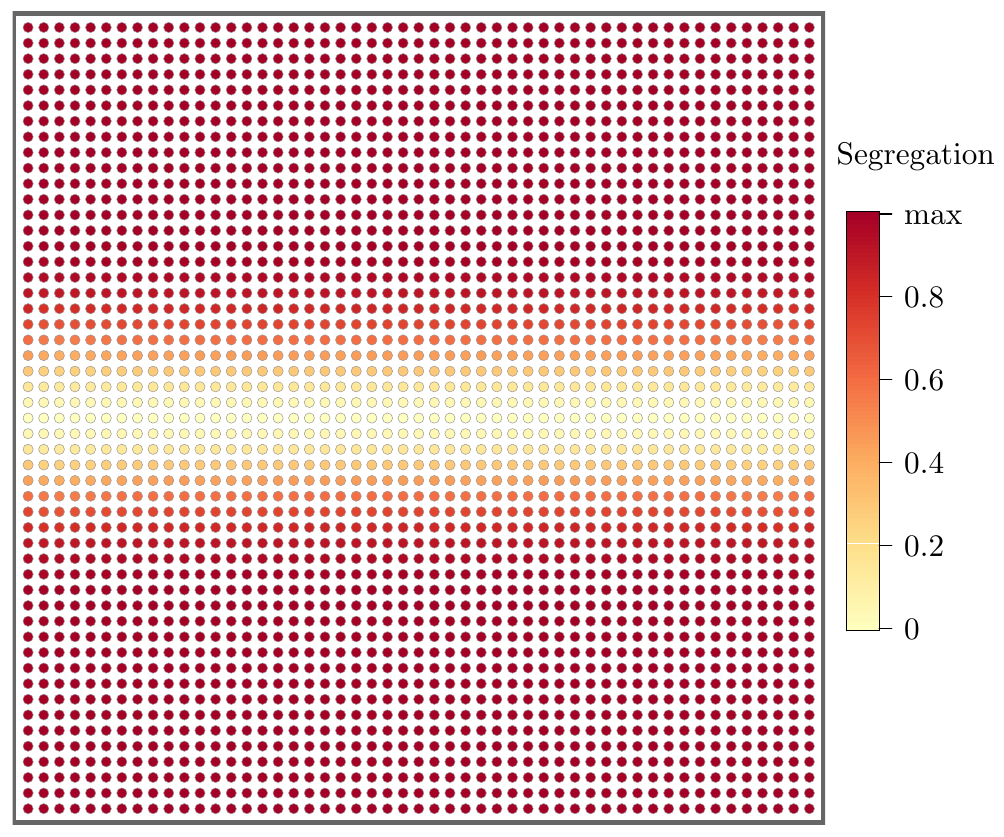}
  \end{subfigure}
  \hspace{1em}
  \begin{subfigure}[t]{0.49\textwidth}
    \caption{Full Barrier}
    \label{fig:half_div2}
    \includegraphics[width=\linewidth, center]{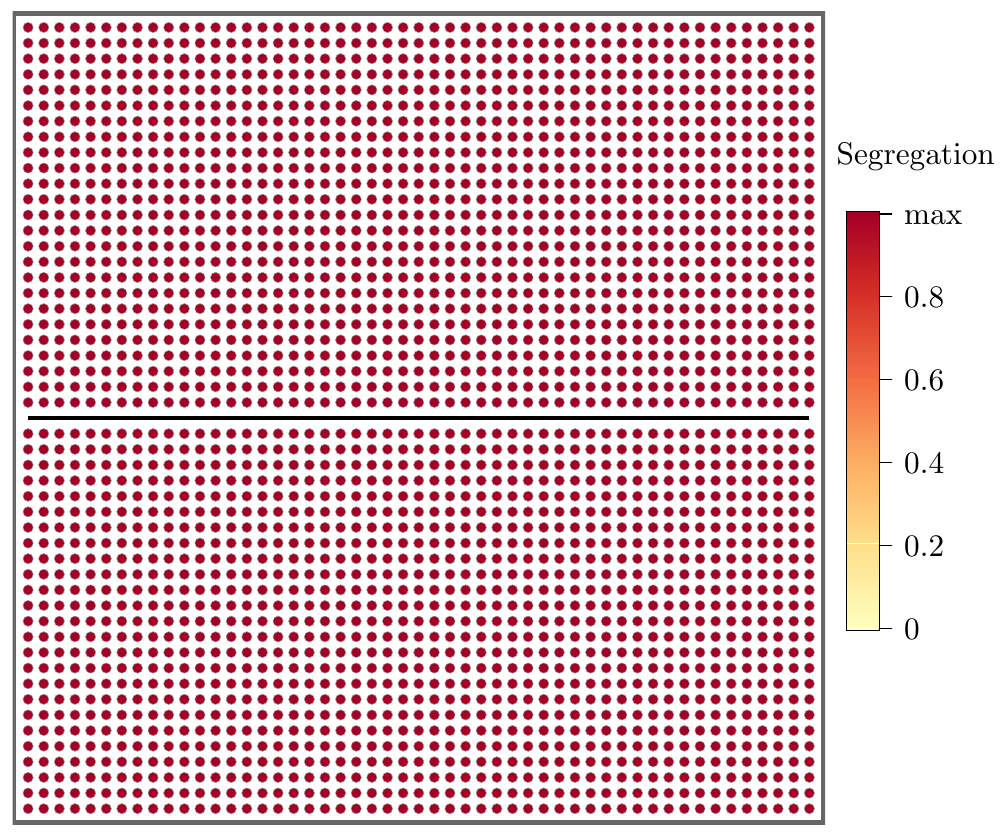}
  \end{subfigure}
  \vspace{2\baselineskip}\\
  \begin{subfigure}[t]{0.49\textwidth}
    \caption{Partial Barrier}
    \label{fig:half_div3}
    \includegraphics[width=\linewidth, center]{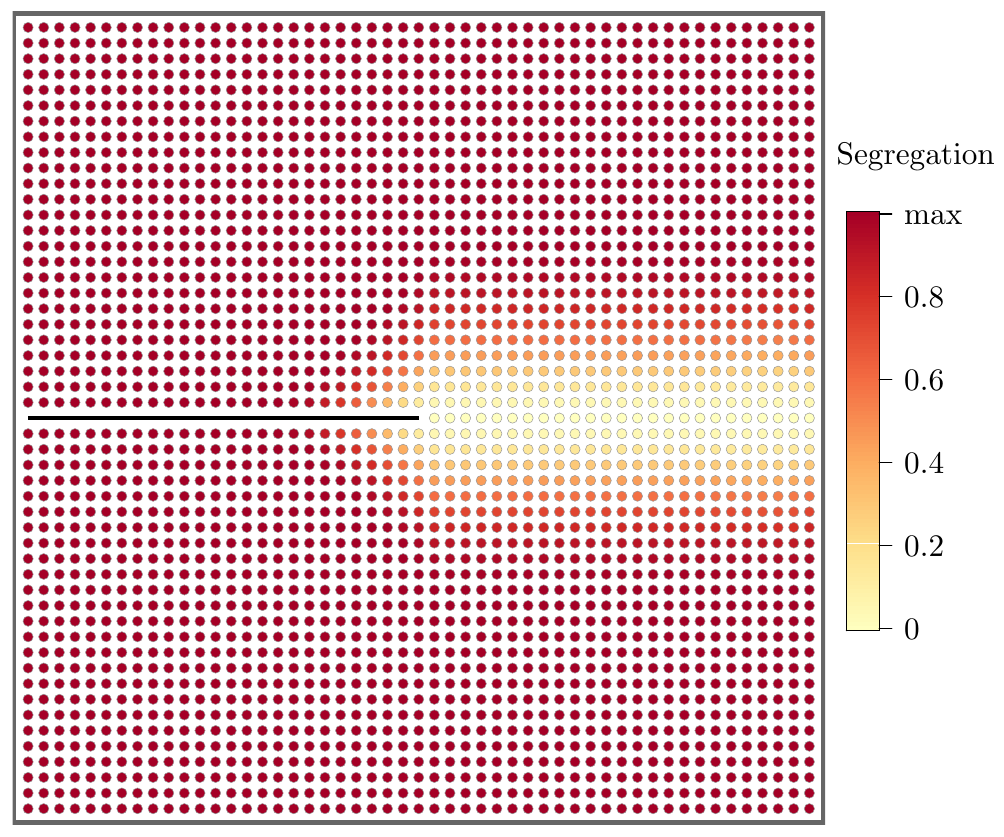}
  \end{subfigure}
  \hspace{1em}
  \begin{subfigure}[t]{0.49\textwidth}
    \caption{Segmented Barrier}
    \label{fig:half_div7}
    \includegraphics[width=\linewidth, center]{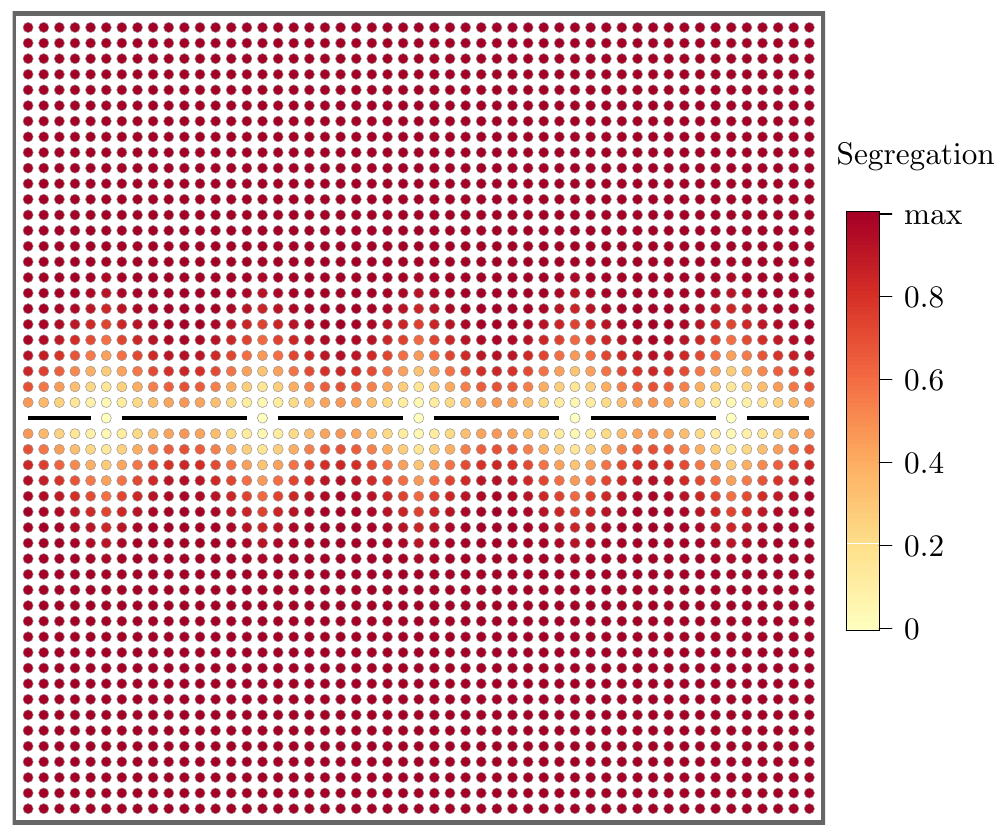}
  \end{subfigure}
  
  \caption{Maps of Local Segregation in a Stylized City by Barrier Type\\ (Reach of Local Environments = 3 km)}
  \label{fig:half_3k}
\end{figure}

\begin{figure}

  \begin{subfigure}[t]{0.49\textwidth}
    \caption{No Barrier}
    \label{fig:half_div0_10k}
    \includegraphics[width=\linewidth, center]{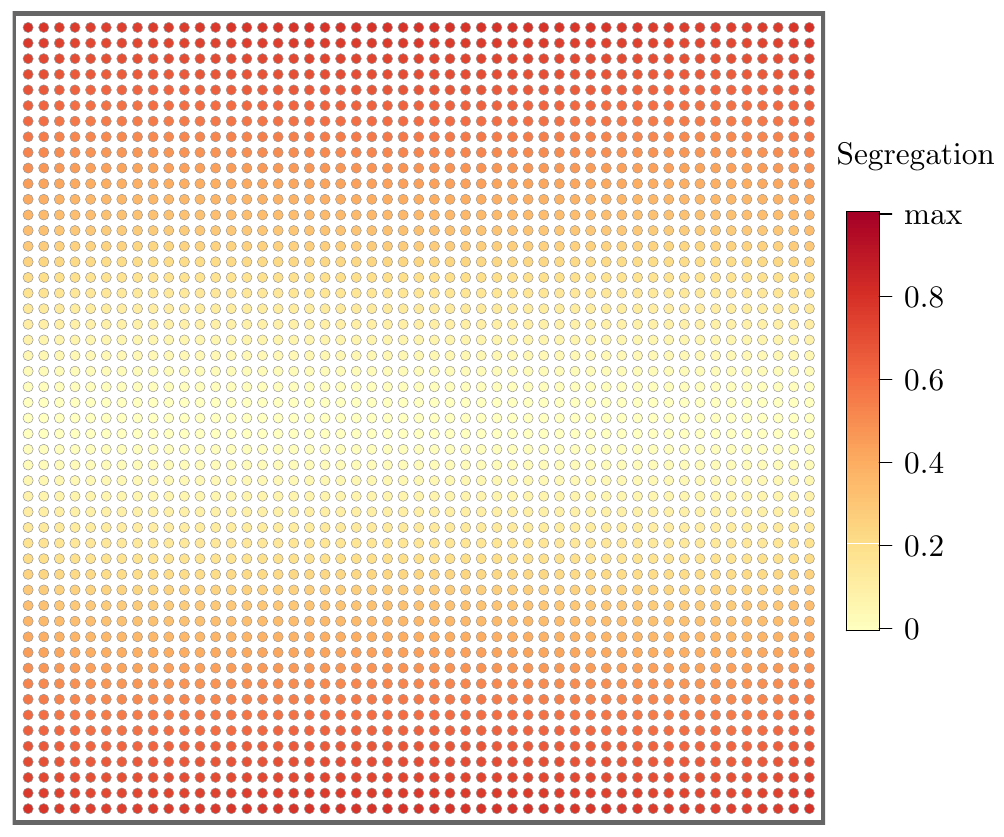}
  \end{subfigure}
  \hspace{1em}
  \begin{subfigure}[t]{0.49\textwidth}
    \caption{Full Barrier}
    \label{fig:half_div2_10k}
    \includegraphics[width=\linewidth, center]{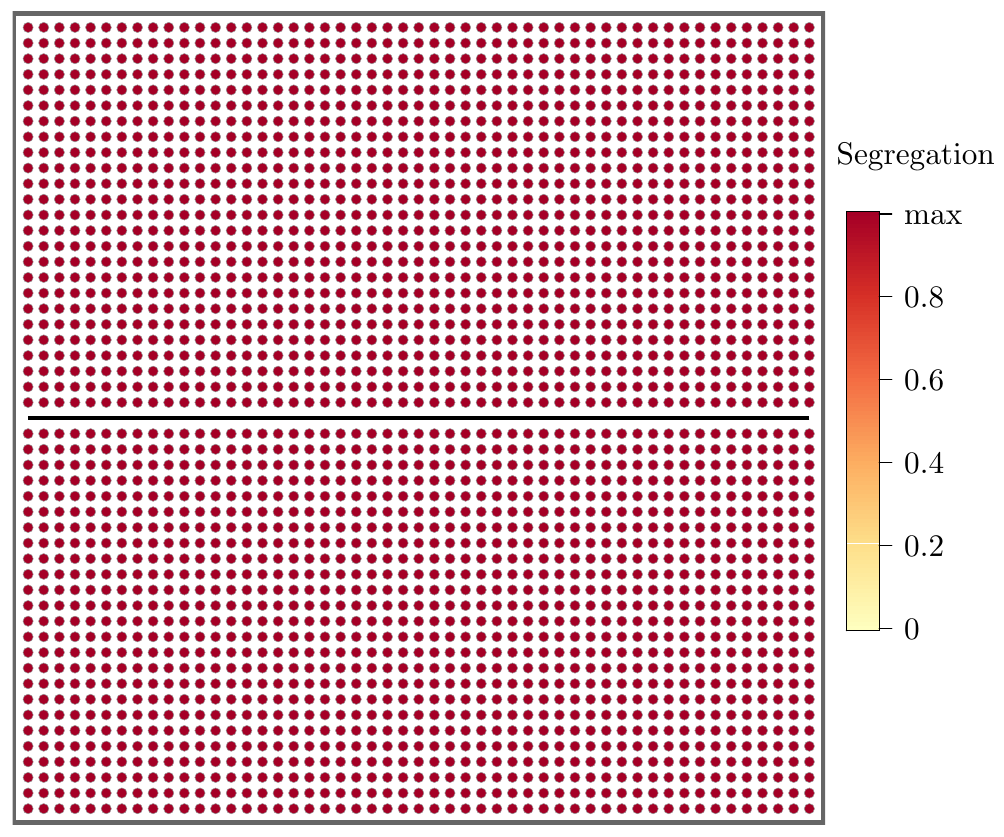}
  \end{subfigure}
  \vspace{2\baselineskip}\\
  \begin{subfigure}[t]{0.49\textwidth}
    \caption{Partial Barrier}
    \label{fig:half_div3_10k}
    \includegraphics[width=\linewidth, center]{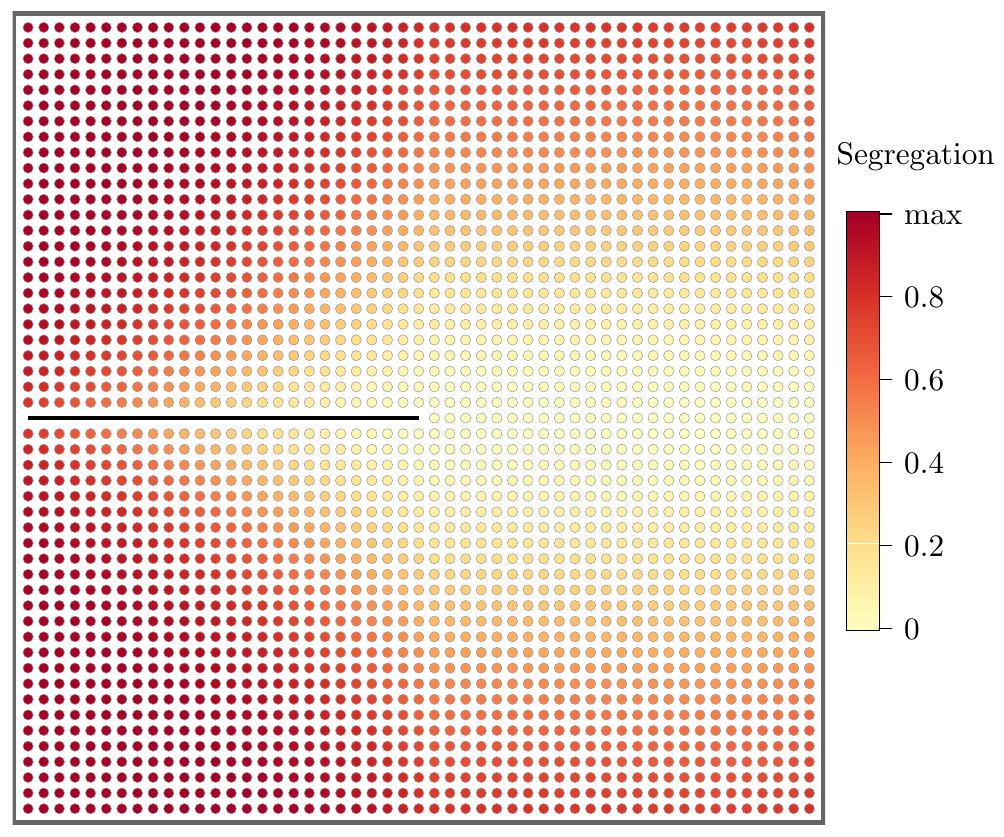}
  \end{subfigure}
   \hspace{1em}
  \begin{subfigure}[t]{0.49\textwidth}
    \caption{Segmented Barrier}
    \label{fig:half_div7_10k}
    \includegraphics[width=\linewidth, center]{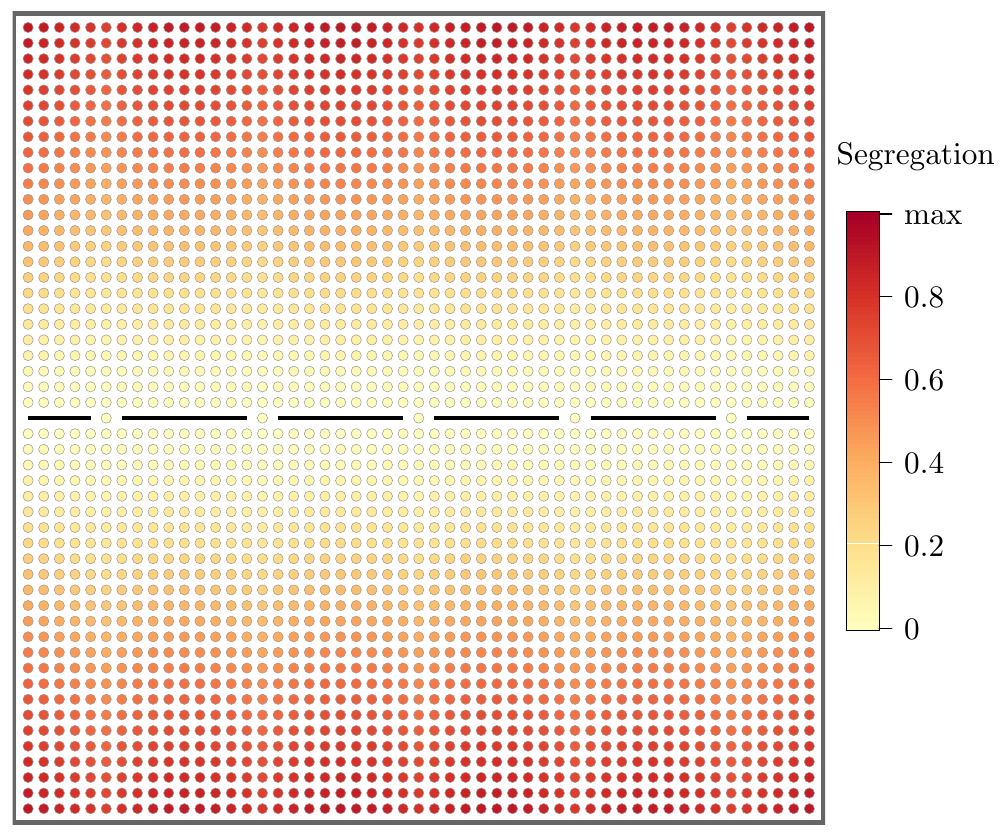}
  \end{subfigure}
  
  \caption{Maps of Local Segregation in a Stylized City by Barrier Type\\ (Reach of Local Environments = 10 km)}
  \label{fig:half_10k}
\end{figure}

The segregation results for the stylized city are summarized in Figure
\ref{fig:ResultsD_half}. Figures \ref{fig:half_3k} and
\ref{fig:half_10k} map the local segregation by barrier type for each
node in the city with local environments that have a reach of 3 km and
10 km, respectively. Darker colors indicate higher segregation.

\begin{center}
[Figure \ref{fig:SegD_half} about here.]
\end{center}

\begin{center}
[Figure \ref{fig:half_3k} about here.]
\end{center}

\begin{center}
[Figure \ref{fig:half_10k} about here.]
\end{center}

Figure \ref{fig:half_div0} maps the results for the city with no
barriers. Segregation values are highest at the north and south end of
the city and lowest along the midline of the city where the two
segregated clusters meet. The city is 12.5 by 12.5 km in size. At a
reach of 3 km, only locations relatively near the midline experience any
change in the composition of their environments.

With no barriers present, the level of segregation is similar using
either distance measure (see Figure \ref{fig:SegD_half}). However, the
measure using straight line distance shows slightly less segregation,
and the difference grows as the reach of local environments increases.
The distance between each pair of locations along the road network is
longer than the the length of a straight line connecting the locations.
Given the same reach, local environments are larger when constructed
with straight line distance than with road distance. They encompass a
larger area of the city and include more of the city's population, which
makes their composition more representative of the city's overall
population.

The small differences between the two sets of results is due to street
design. If the city's simple street grid was amended to include streets
running diagonally through each block, local environments constructed
with either distance measure would include the same set of locations,
and segregation results would be identical. For example, in cities with
avenues that periodically cross diagonally through a grid of streets, as
in parts of Washington, DC, the straight line distance and road distance
between two locations is more similar than in cities with a rectangular
street grid, as in Midtown Manhattan. The presence of dead end streets
and cul-de-sacs has the opposite effect on connectivity: they prevent
through movement and increase the difference between the straight line
distance and road distance between locations.

Now that I have established the differences in the straight line
distance and road distance segregation measures that are attributable to
the stylized city's street design, I separately introduce each of the
three barriers into the city and again measure segregation and compare
results.

The first barrier fully disconnects the city's two large clusters and
divides the north and south sides of the city (see Figure
\ref{fig:half_div2}). As a consequence, local environments constructed
with road distance do not include areas on the opposite side of the
barrier. Segregation remains at its maximum value, even as the reach of
local environments increases. Figure \ref{fig:CompareD_half} shows the
difference between the straight line distance and road distance
segregation measures for each of the barrier types. Road distance is
sensitive to the disconnection imposed by the barrier, but straight line
distance is not. Local environments constructed with straight line
distance are unchanged by the presence of a barrier. There is maximum
segregation in the immediate area of each location, but segregation
steadily decreases as the reach of local environments increases.

The full barrier in Figure \ref{fig:half_div2} is an extreme case that
is rarely observed at the city level. Figures \ref{fig:half_div3} and
\ref{fig:half_div7} show the effect of more realistic partial barriers.
The barrier in Figure \ref{fig:half_div3} spans half the city. This
creates excess distance between locations on either side of the barrier,
but no locations are fully disconnected from the rest of the city.
Similarly, in Figure \ref{fig:half_div7} there is a segmented barrier
that creates excess distance, but maintains connectivity, similar to a
river with bridges every few kilometers. The straight line distance
segregation measures show the same results, regardless of whether one of
the barriers is present. The effect of the barriers is only evident when
using road distance segregation measures.

The presence of a segmented barrier results in higher segregation,
compared to a city with no barrier. Segregation decreases slowly as
local environments increase to a reach of 3 km, and then shows a steeper
rate of decline beyond that distance. (See Figure
\ref{fig:ResultsD_half}.) The difference narrows when the reach is 10 km
and nearly converges with the segregation level for a city with no
barrier.

With a barrier that spans half the city's width, segregation decreases
as the reach of local environments increases. The rate of change is
steady, but it is more gradual than when no barrier is present. As the
reach of local environments increases, the difference in segregation
between the city with no barrier and a barrier that spans half the city
also increases (see Figure \ref{fig:ResultsD_half}). This is opposite to
the trend observed for the city with a segmented barrier where the
difference narrowed. This difference occurs because the barriers
constrain connectivity in different ways.

Both barriers create excess distance between locations, but the total
length of the segmented barrier in Figure \ref{fig:half_div7} is greater
than the barrier that spans half the city in Figure \ref{fig:half_div3}.
In addition, the segmented barrier is more permeable, allowing
connectivity between the two halves of the city at regular intervals.
Although the segmented barrier facilitates higher segregation when the
reach of local environments is less than 3 km, it is less impactful at
greater distances. However, the barrier that spans half the city
continues to constrain the local environments of much of the city's
population, even when the reach is 10 km. The difference in the impact
of each barrier is evident in Figure \ref{fig:half_10k}, which maps the
segregation by barrier type for local environments with a reach of 10
km.

\subsubsection{\texorpdfstring{``The Other Side of the Tracks'' in
Pittsburgh,
PA}{The Other Side of the Tracks in Pittsburgh, PA}}\label{the-other-side-of-the-tracks-in-pittsburgh-pa}

To further demonstrate the application of the SPC method, I examine the
impact of physical barriers on residential segregation levels in a U.S.
city: Pittsburgh, PA. I measure racial segregation between whites,
blacks, Hispanics, and Asians using data from the 2010 decennial
census.\footnote{Using U.S. Census Bureau's categories of race and
  ethnicity, I define four ethnoracial groups: non-Hispanic white
  (``white''), non-Hispanic black (``black''), non-Hispanic Asian
  (``Asian''), and Hispanic of any race (``Hispanic'').} I measure
segregation for local environments with a reach of .5 km, 1 km, 2 km, 3
km, and 4 km. The smallest reach of .5 km approximates the area of a
neighborhood in Pittsburgh, and the largest reach of 4 km would include
a large portion of the city. I compare the road distance and straight
line distance segregation measures for each reach to examine how
physical barriers influence the segregation levels.

\begin{figure}
  \includegraphics[height=4.5in, center]{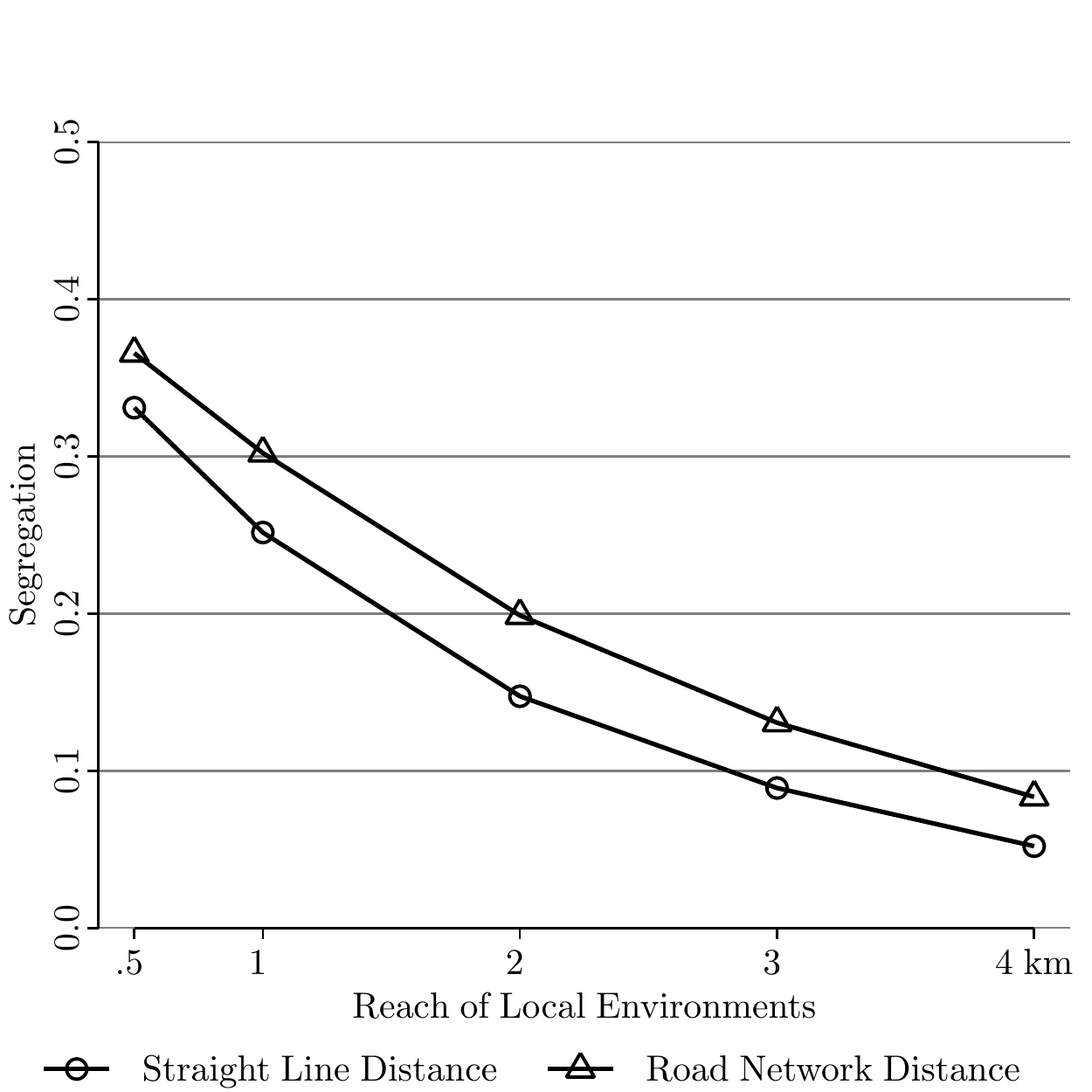}
  \caption{White-Black-Hispanic-Asian Segregation in Pittsburgh in 2010\\ Comparison of Road Distance and Straight Line Distance Segregation Measures}
  \label{fig:pt_segbydist_line}
\end{figure}

The population of Pittsburgh is 65 percent white, 26 percent black, 4
percent Asian, and 2 percent Hispanic. However, local environments
within the city tend to have a different composition than the city.
Figure \ref{fig:pt_segbydist_line} presents the results for the road
distance and straight line distance segregation measures. The city-level
segregation for local environments with a reach of .5 km is .33 for the
straight line distance segregation measure and .37 for the road distance
segregation measure. Both segregation measures steadily decrease as the
reach of local environments increases, but the level of segregation is
consistently higher for the road distance segregation measure.

\begin{center}
[Figure \ref{fig:pt_segbydist_line} about here.]
\end{center}

The magnitude of the difference between the straight line distance and
road distance measures indicates how much of an impact physical barriers
have on segregation. However, the city-level segregation results for
each reach are the population-weighted average of segregation in the
local environments of each node, including areas where there are no
physical barriers. In such areas, the local environments constructed
with road distance and straight line distance will have a similar
composition and there will no difference between the two segregation
measures. Therefore, even small positive differences in the city-level
results are meaningful and suggest that physical barriers facilitate
greater separation between ethnoracial groups and higher levels of
segregation.

\begin{figure}
  \includegraphics[width=\textwidth, center]{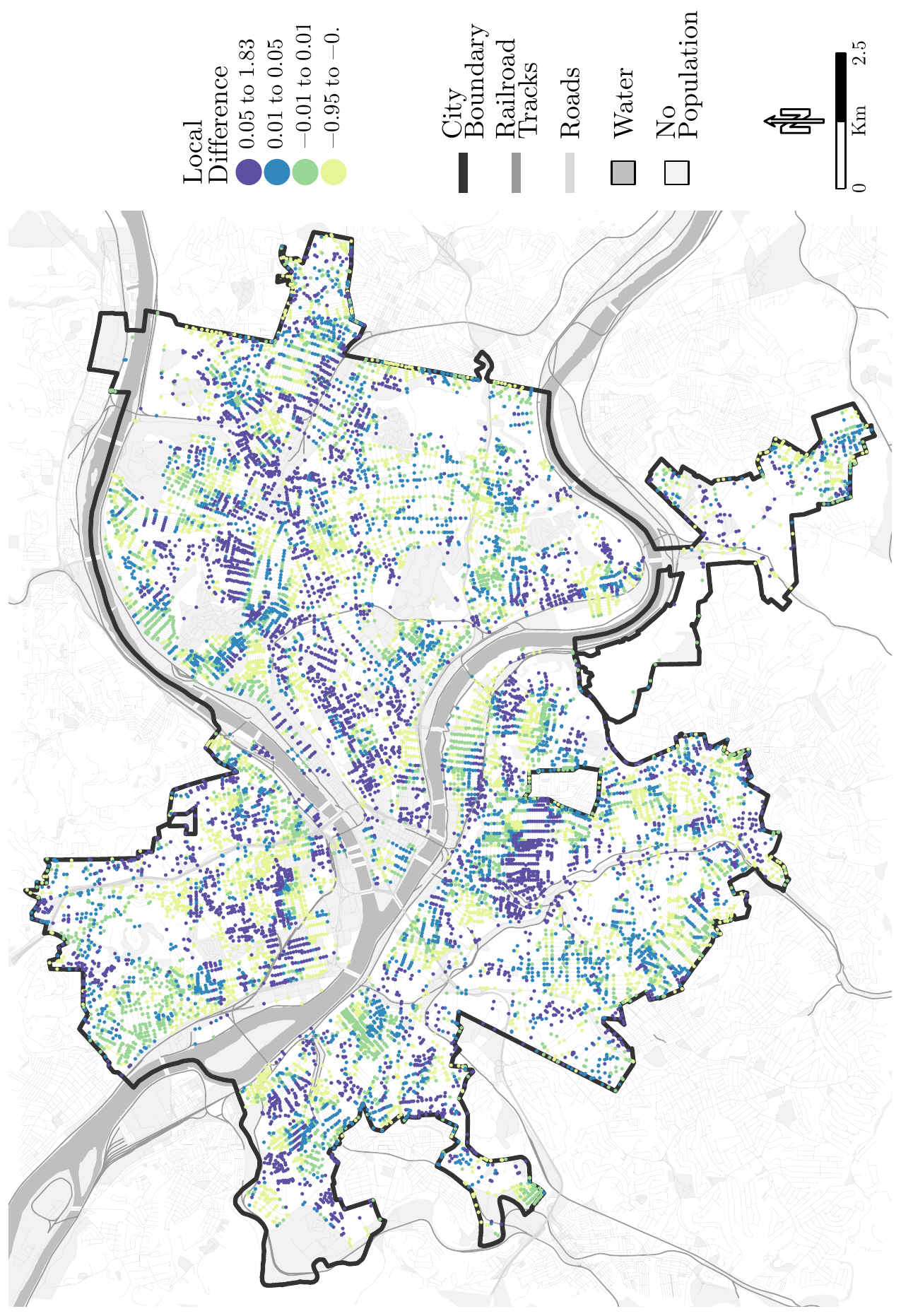}
  \caption{White-Black-Hispanic-Asian Segregation in Pittsburgh in 2010\\ Quartiles of the Difference between Road Distance and Straight Line Distance Segregation Measures\\ (Reach of Local Environments = .5 km)}
  \label{fig:pt_quart_500m}
\end{figure}

The difference between the road distance and straight line distance
segregation measures varies considerably across locations in the city.
Figure \ref{fig:pt_quart_500m} is a map of the differences in local
environments with a reach of .5 km. The differences are grouped into
quartiles and the color of each node indicates the magnitude of the
difference, with darker colors indicating a greater difference between
the measures. There are areas where a cluster of locations all have
larger differences between the measures, which indicates that road
connectivity or physical barriers in these areas is facilitating higher
levels of segregation. A closer inspection of one of these areas
illustrates how such differences can arise.

\begin{center}
[Figure \ref{fig:pt_quart_500m} about here.]
\end{center}

\begin{figure}
  \includegraphics[height=3.5in, center]{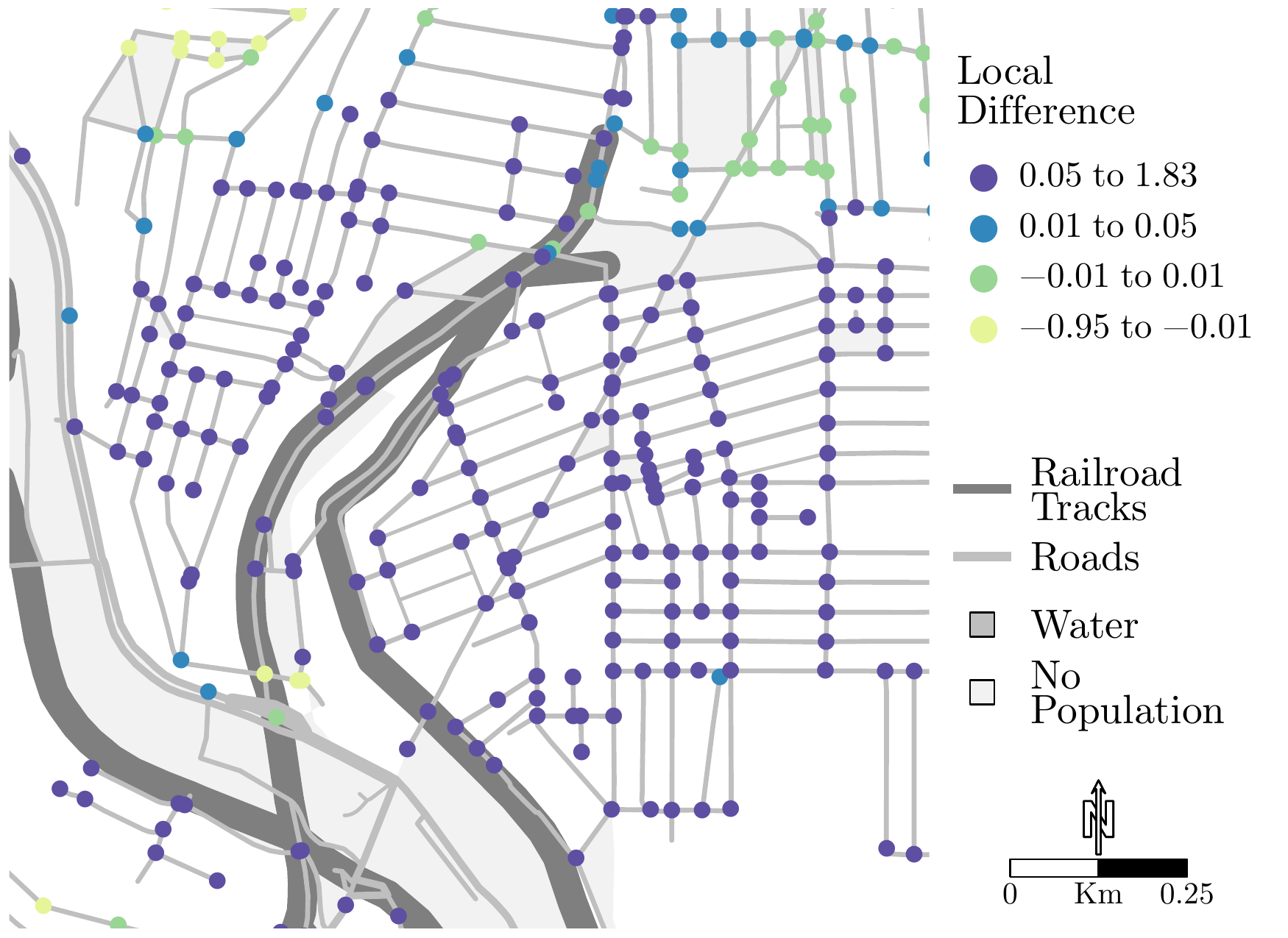}
  \caption{White-Black-Hispanic-Asian Segregation in an area of Pittsburgh in 2010\\ Quartiles of the Difference between Road Distance and Straight Line Distance Segregation Measures\\ (Reach of Local Environments = .5 km)}
  \label{fig:pt_quart_500m_zoom}
\end{figure}

Figure \ref{fig:pt_quart_500m_zoom} is a map of the local differences
between the road distance and straight line distance segregation
measures in the Beltzhoover neighborhood of Pittsburgh and other nearby
areas. Beltzhoover is located south of downtown Pittsburgh in the
Hilltop area of the city. It is is bounded on the west and north by
railroad tracks and on the south by McKinley Park. These features of the
built environment reduce the connectivity and increase the distance
between Beltzhoover and the neighborhoods of Mount Washington to the
west and north and Bon Air to the south. The road network distance
between locations in Beltzhoover and these nearby neighborhoods is
longer than a straight line connecting the locations. Due to this
difference, the local environments of residents in these neighborhoods
will differ depending on whether their reach is measured with straight
line distance or road distance.

\begin{center}
[Figure \ref{fig:pt_quart_500m_zoom} about here.]
\end{center}

For smaller reaches, such as .5 km, the local environments of locations
near the railroad tracks will include areas on opposite sides of the
tracks if they are constructed with straight line distance, but not if
they are constructed with road distance. Figure \ref{fig:rle_pt}, used
as an example earlier in the paper, illustrates this difference. A reach
of .5 km is not a sufficient distance to connect the focal location in
Beltzhoover to locations on the other side of the tracks in Mount
Washington along the road network.

\begin{figure}
  \includegraphics[height=3.5in, center]{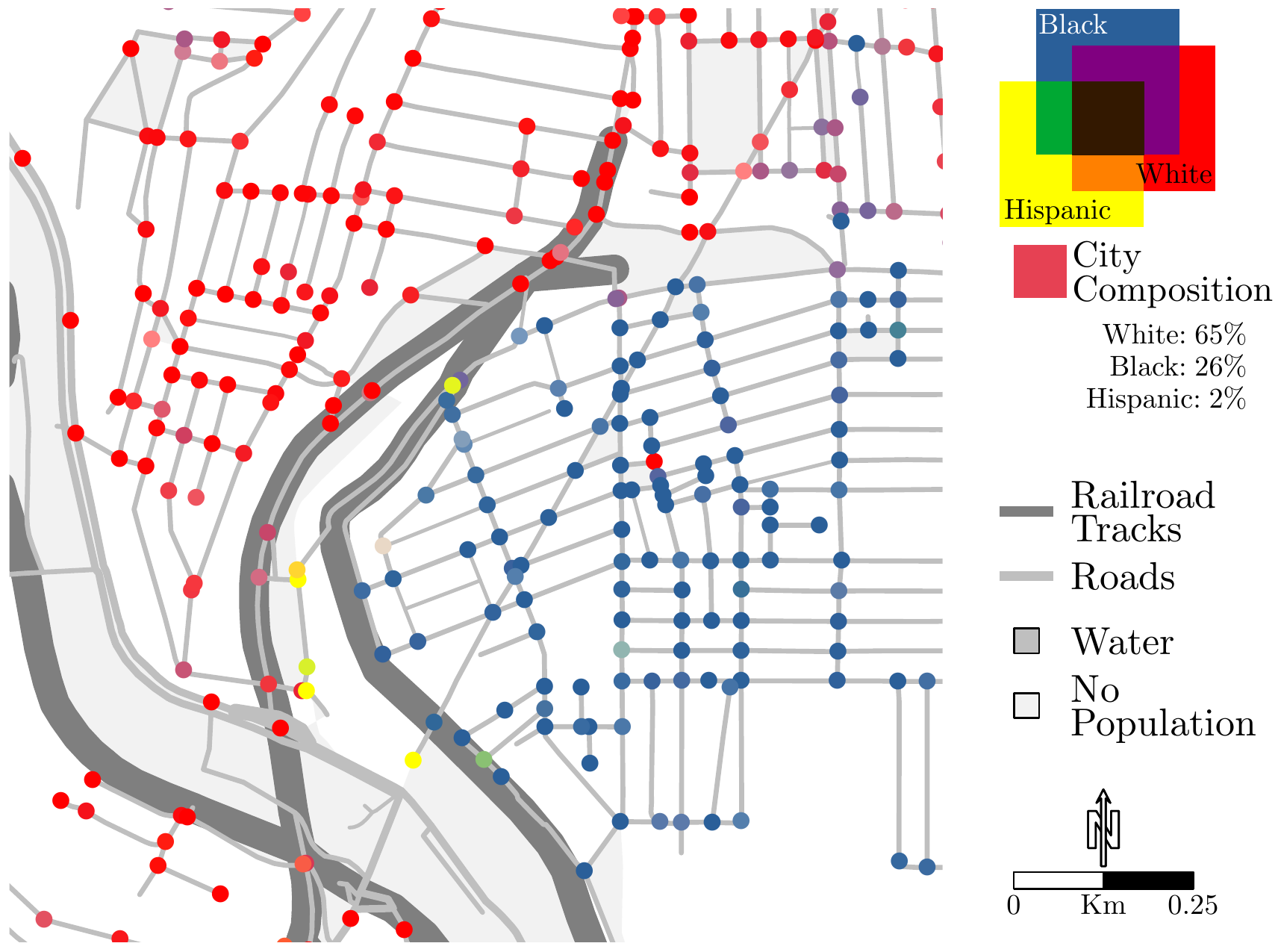}
  \caption{White, Black, and Hispanic Population in an area of Pittsburgh in 2010}
  \label{fig:pt_wbh_zoom}
\end{figure}

The neighborhoods of Beltzhoover and Mount Washington are physical
divided by the railroad tracks, and they also differ in their racial
composition. Figure \ref{fig:pt_wbh_zoom} is a map of the
black-Hispanic-white composition in this area of Pittsburgh. The colored
box labeled ``City Composition'' indicates the color that a node will be
if it matches the black-Hispanic-white composition of the city. Most
residents on the Beltzhoover side of the tracks are black and most
residents on the Mount Washington side are white. The city of Pittsburgh
is 65 percent white and 26 percent black, which is different than the
predominantly black composition of Beltzhoover.

\begin{center}
[Figure \ref{fig:pt_wbh_zoom} about here.]
\end{center}

\begin{figure}

  \begin{subfigure}[t]{\textwidth}
    \caption{Straight Line Distance Segregation Measure}
    \label{fig:pt_ed_500m}
    \includegraphics[height=3.5in, center]{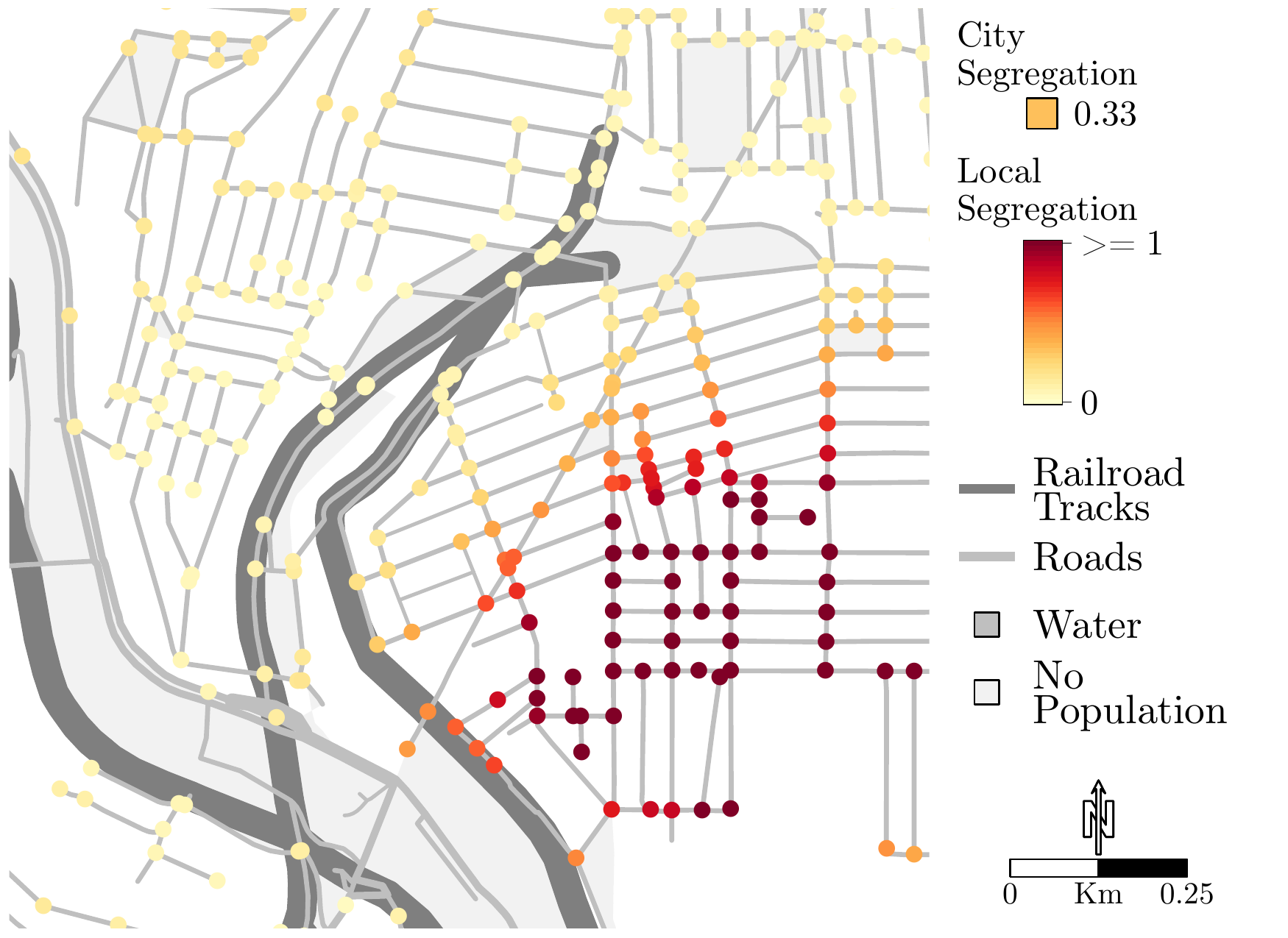}
  \end{subfigure}
  ~\\
  \begin{subfigure}[t]{\textwidth}
    \caption{Road Distance Segregation Measure}
    \label{fig:pt_pd_500m}
    \includegraphics[height=3.5in, center]{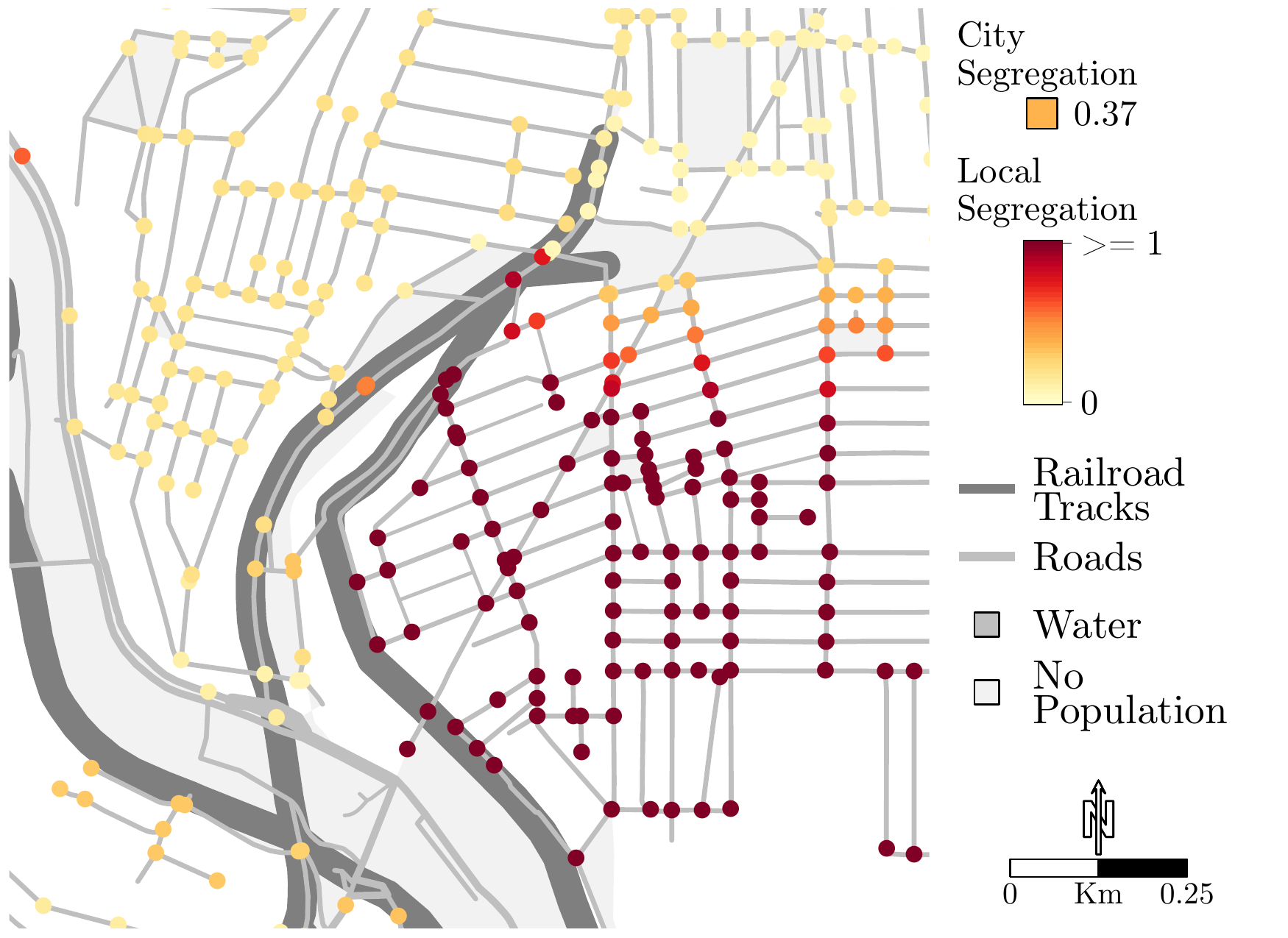}
  \end{subfigure}

  \caption{White-Black-Hispanic-Asian Segregation in an area of Pittsburgh in 2010\\ Straight Line Distance and Road Distance Segregation Measures\\ (Reach of Local Environments = .5 km)}
  \label{fig:pt_segbydist_maps}
\end{figure}

Because the racial composition of Beltzhoover is surprising given the
overall composition of the city's population, we should expect
segregation to be high in the local environments of locations in
Beltzhoover, particularly for smaller reaches measured with road
distance. Figure \ref{fig:pt_segbydist_maps} maps the local segregation
values for local environments with a .5km reach, with darker colors
indicating higher segregation values. The results for the straight line
distance segregation measure are shown in Figure \ref{fig:pt_ed_500m}
and the results for the road distance segregation measure are in Figure
\ref{fig:pt_pd_500m}.

Comparing the two maps in Figure \ref{fig:pt_segbydist_maps} reveals the
extent to which the built environment, including road connectivity and
physical barriers, impacts the segregation of these local environments.
The segregation values for locations near the railroad tracks in
Beltzhoover are higher when their local environments are constructed
with road distance rather than straight line distance. The local
environments constructed with straight line distance are able to extend
into the Mount Washington neighborhood, as illustrated in Figure
\ref{fig:rle_pt}. In doing so, their racial composition is more
representative of the city, and therefore less segregated, than the
local environments constructed with road distance.

\begin{center}
[Figure \ref{fig:pt_segbydist_maps} about here.]
\end{center}

Measuring distance along a city's road network is sensitive to the
reduced connectivity and excess distance created by physical barriers.
Comparing road distance and straight line distance segregation measures
reveals the extent to which these physical barriers facilitate greater
separation between groups and increase the local and city-wide levels of
segregation.

\section{Conclusion}\label{conclusion}

The SPC method measures segregation as a function of road network
distance to capture how the physical structure of the built environment
affects the proximity and connectivity of residential locations. I
demonstrated an application of the SPC method with two analyses that
compared road distance segregation measures and straight line distance
segregation measures to examine the impact of physical barriers on
residential segregation. The first analysis compared how segregation
differs in a stylized city when various types of physical barriers
spatially separate two groups, and it showed that a barrier's impact on
segregation depends on its spatial configuration and how much it
restricts the connectivity between groups. In the second analysis of
racial segregation in Pittsburgh, I found that physical barriers divide
urban space in ways that increase the city's overall level of
segregation. However, I also found substantial variation in the impact
of barriers across local areas within the city. It is likely that the
prevalence of barriers and their impact on segregation levels varies
both within cities and across cities as well.

My approach contributes to the scholarship on residential segregation by
capturing the effect of an additional mechanism of segregation---the
connectivity, or physical barriers, between locations---on the level and
spatial pattern of segregation. It uncovers an additional source of
variation in the segregation experienced by city residents, which has
important implications for understanding the causes and consequences of
segregation. Although I have demonstrated how the method can be used to
measure racial residential segregation, the SPC method approach can be
used to measure the spatial segregation or inequality of any population
or environmental characteristic, such as income, crime, or environmental
hazards.

While this approach improves upon the current methods for measuring
spatial segregation, there are two issues that present challenges to its
implementation. First, the method requires detailed spatial data, which
are only available in digital format for the most recent census years. A
historical analysis of segregation patterns using SPC would require the
collection and digitization of street maps and block boundaries from
previous census years. Second the spatially detailed measurement and
analysis required by the SPC method is computationally-intensive. The
processing, memory, and storage demands of the method may not be
feasible for many personal computers. For example, the road network data
for the city of Pittsburgh includes approximately 19,000 nodes. Thus,
calculating the road network distance requires finding the shortest path
between 186 million pairs of nodes and storing the result in a large
distance matrix. An additional matrix containing proximity weights is
computed for every reach of the local environments. The 12 matrices used
in the demonstration analysis required about 35 gigabytes of data
storage and the measurement and analysis was conducted in a high
performance computing environment. Given the rapid rate of advances in
computational resources, including the availability of high performance
computing facilities and personal computers with multiple core
processors and large memory and storage capacities, the computational
demands of the method will become increasingly easy to meet.

Despite these limitations, the SPC method can be extended in a variety
of ways to incorporate additional aspects of proximity and connectivity.
It is possible to study how the connectivity provided by public
transportation segregates groups by using a network of transit routes
and stations instead of a road network, or by integrating the
connectivity provided by both transit and roads into a single network.
Or, the road network can be constructed as a directed graph if, for
example, one wanted to represent the differences in connectivity
provided by one-way vs.~two-way streets. Further, the method can
incorporate additional information about the quality of the connectivity
provided by roads and other pathways. For example, a bridge or underpass
that is desolate or poorly lit may perpetuate separation between nearby
areas rather than bringing them together. A measure of the quality of
connective roads could further enhance the SPC approach, and new methods
for systematic social observation using Google Street View could provide
a starting point for developing such a measure (Hwang and Sampson 2014).

The SPC method can also accommodate alternative theories of proximity.
If one was interested in measuring segregation as a function of
population count (i.e., to control for differences in population density
across cities or neighborhoods), the reach of the local environments can
be based on the \emph{count} of nearest neighbors rather than the
\emph{distance} between nodes. Or, if the concern of a study was spatial
\emph{mobility} rather than spatial \emph{structure}, the travel time
between nodes can be used to define the reach of local environments
instead of road network distance. Such an extension of the SPC method
may be relevant to research that uses an activity space perspective to
study segregation across the many social and geographic spaces where
individuals travel and spend time on a day-to-day basis (Jones and
Pebley 2014; Schnell and Yoav 2001; Wong and Shaw 2011; Zenk et al.
2011), or related research that uses geo-ethnography to study the
activity patterns that link people to places (Matthews 2011).

I designed the SPC method to create a more accurate and comprehensive
portrait of the physical environment of individuals' residential spaces
and develop a deeper understanding of how the built environment
influences the patterns, processes, and consequences of segregation. My
approach highlights the unique role of the built environment in
spatially structuring segregation patterns and enables further
consideration its role in the persistence of segregation.

\clearpage

\section{References}\label{references}

\setlength{\parskip}{1ex plus 0ex} \setlength{\leftskip}{2em}
\setlength{\parindent}{-2em} \indent

\hyperdef{}{ref-Ananat:2011jc}{\label{ref-Ananat:2011jc}}
Ananat, Elizabeth Oltmans. 2011. ``The Wrong Side(s) of the Tracks: The
Causal Effects of Racial Segregation on Urban Poverty and Inequality.''
\emph{American Economic Journal: Applied Economics} 3(2):34--66.

\hyperdef{}{ref-Anderson:1990us}{\label{ref-Anderson:1990us}}
Anderson, Elijah. 1990. \emph{Streetwise: Race, Class, and Change in an
Urban Community}. Chicago : University of Chicago Press.

\hyperdef{}{ref-Atkinson:2004es}{\label{ref-Atkinson:2004es}}
Atkinson, Rowland and John Flint. 2004. ``Fortress UK? Gated
communities, the spatial revolt of the elites and time-space
trajectories of segregation.'' \emph{Housing Studies} 19(6):875--92.

\hyperdef{}{ref-Bader:2015dq}{\label{ref-Bader:2015dq}}
Bader, Michael D. M. and Maria Krysan. 2015. ``Community Attraction and
Avoidance in Chicago: What's Race Got to Do with It?'' \emph{Annals of
the American Academy of Political and Social Science} 660(1):261--81.

\hyperdef{}{ref-Besbris:2015de}{\label{ref-Besbris:2015de}}
Besbris, Max, Jacob William Faber, Peter Rich, and Patrick Sharkey.
2015. ``Effect of Neighborhood Stigma on Economic Transactions.''
\emph{Proceedings of the National Academy of Sciences} 112(16):4994--98.

\hyperdef{}{ref-Bischoff:2008bw}{\label{ref-Bischoff:2008bw}}
Bischoff, Kendra. 2008. ``School District Fragmentation and Racial
Residential Segregation How Do Boundaries Matter?'' \emph{Urban Affairs
Review} 44(2):182--217.

\hyperdef{}{ref-R-rgeos}{\label{ref-R-rgeos}}
Bivand, Roger S. and Colin Rundel. 2014. \emph{rgeos: Interface to
Geometry Engine - Open Source (GEOS)}.

\hyperdef{}{ref-R-rgdal}{\label{ref-R-rgdal}}
Bivand, Roger S., Tim Keitt, and Barry Rowlingson. 2014. \emph{rgdal:
Bindings for the Geospatial Data Abstraction Library}.

\hyperdef{}{ref-Blakely:1997uk}{\label{ref-Blakely:1997uk}}
Blakely, Edward J. and Mary Gail Snyder. 1997. \emph{Fortress America}.
Brookings Institution Press.

\hyperdef{}{ref-Brown:2006hb}{\label{ref-Brown:2006hb}}
Brown, Lawrence A. and Su-Yeul Chung. 2006. ``Spatial segregation,
segregation indices and the geographical perspective.''
\emph{Population, Space and Place} 12(2):125--43.

\hyperdef{}{ref-Bruch:2014es}{\label{ref-Bruch:2014es}}
Bruch, Elizabeth E. 2014. ``How Population Structure Shapes Neighborhood
Segregation.'' \emph{The American Journal of Sociology} 119(5):1221--78.

\hyperdef{}{ref-Charles:2003ca}{\label{ref-Charles:2003ca}}
Charles, Camille Zubrinsky. 2003. ``The Dynamics of Racial Residential
Segregation.'' \emph{Annual Review of Sociology} 29:167--207.

\hyperdef{}{ref-Crowder:2008wx}{\label{ref-Crowder:2008wx}}
Crowder, Kyle and Scott J. South. 2008. ``Spatial Dynamics of White
Flight: The Effects of Local and Extralocal Racial Conditions on
Neighborhood Out-Migration.'' \emph{American Sociological Review}
73(5):792--812.

\hyperdef{}{ref-Csardi:2006vh}{\label{ref-Csardi:2006vh}}
Csardi, Gabor and Tamas Nepusz. 2006. ``The igraph Software Package for
Complex Network Research.'' \emph{InterJournal}.

\hyperdef{}{ref-Duncan:1955ve}{\label{ref-Duncan:1955ve}}
Duncan, Otis Dudley and Beverly Duncan. 1955. ``A Methodological
Analysis of Segregation Indexes.'' \emph{American Sociological Review}
20(2):210--17.

\hyperdef{}{ref-Farrell:2008hh}{\label{ref-Farrell:2008hh}}
Farrell, Chad R. 2008. ``Bifurcation, Fragmentation or Integration? The
Racial and Geographical Structure of US Metropolitan Segregation,
1990--2000.'' \emph{Urban Studies} 45(3):467--99.

\hyperdef{}{ref-Fischer:2004tk}{\label{ref-Fischer:2004tk}}
Fischer, Claude S., Gretchen Stockmayer, Jon Stiles, and Michael Hout.
2004. ``Distinguishing the Geographic Levels and Social Dimensions of
U.S. Metropolitan Segregation, 1960-2000.'' \emph{Demography}
41(1):37--59.

\hyperdef{}{ref-Fischer:2008kx}{\label{ref-Fischer:2008kx}}
Fischer, Mary J. 2008. ``Shifting Geographies: Examining the Role of
Suburbanization in Blacks' Declining Segregation.'' \emph{Urban Affairs
Review} 43(4):475--96.

\hyperdef{}{ref-Folch:2016cf}{\label{ref-Folch:2016cf}}
Folch, David C. and Sergio J. Rey. 2016. ``The Centralization Index: a
Measure of Local Spatial Segregation.'' \emph{Papers in Regional
Science} 95(3):555.

\hyperdef{}{ref-Fotheringham:1991ur}{\label{ref-Fotheringham:1991ur}}
Fotheringham, A. Stewart and David W. S. Wong. 1991. ``The Modifiable
Areal Unit Problem in Multivariate Statistical-Analysis.''
\emph{Environment and Planning A} 23(7):1025--44.

\hyperdef{}{ref-Fowler:2015ce}{\label{ref-Fowler:2015ce}}
Fowler, Christopher S. 2015. ``Segregation as a multiscalar phenomenon
and its implications for neighborhood-scale research: the case of South
Seattle 1990--2010.'' \emph{Urban Geography} 37(1):1--25.

\hyperdef{}{ref-Fowler:2016bk}{\label{ref-Fowler:2016bk}}
Fowler, Christopher S., Barrett A. Lee, and Stephen A. Matthews. 2016.
\emph{Demography} 1--23.

\hyperdef{}{ref-Grannis:1998ui}{\label{ref-Grannis:1998ui}}
Grannis, Rick. 1998. ``The Importance of Trivial Streets: Residential
Streets and Residential Segregation.'' \emph{American Journal of
Sociology} 103(6):1530--64.

\hyperdef{}{ref-Grigoryeva:2015bm}{\label{ref-Grigoryeva:2015bm}}
Grigoryeva, Angelina and Martin Ruef. 2015. ``The Historical Demography
of Racial Segregation.'' \emph{American Sociological Review}
80(4):814--42.

\hyperdef{}{ref-Hunter:1974tl}{\label{ref-Hunter:1974tl}}
Hunter, Albert. 1974. \emph{Symbolic Communities: The Persistence and
Change of Chicago's Local Communities}. Chicago : University of Chicago
Press.

\hyperdef{}{ref-Hwang:2016kc}{\label{ref-Hwang:2016kc}}
Hwang, Jackelyn. 2016. ``The Social Construction of a Gentrifying
Neighborhood: Reifying and Redefining Identity and Boundaries in
Inequality.'' \emph{Urban Affairs Review} 52(1):98--128.

\hyperdef{}{ref-Hwang:2014ku}{\label{ref-Hwang:2014ku}}
Hwang, Jackelyn and Robert J. Sampson. 2014. ``Divergent Pathways of
Gentrification.'' \emph{American Sociological Review} 79(4):726--51.

\hyperdef{}{ref-Jackson:1985wr}{\label{ref-Jackson:1985wr}}
Jackson, Kenneth T. 1985. \emph{Crabgrass Frontier: The Suburbanization
of the United States}. New York: Oxford University Press.

\hyperdef{}{ref-Jargowsky:2005tx}{\label{ref-Jargowsky:2005tx}}
Jargowsky, Paul A. and J. Kim. 2005. ``A Measure of Spatial Segregation:
the Generalized Neighborhood Sorting Index.'' \emph{University of Texas,
Dallas}.

\hyperdef{}{ref-Jones:2014cn}{\label{ref-Jones:2014cn}}
Jones, Malia and Anne R. Pebley. 2014. ``Redefining Neighborhoods Using
Common Destinations: Social Characteristics of Activity Spaces and Home
Census Tracts Compared.'' \emph{Demography} 51(3):727--52.

\hyperdef{}{ref-R-bigmemory}{\label{ref-R-bigmemory}}
Kane, Michael J., John W. Emerson, and Peter Haverty. 2013.
\emph{bigmemory: Manage massive matrices with shared memory and
memory-mapped files}.

\hyperdef{}{ref-RoryKramer:2017kc}{\label{ref-RoryKramer:2017kc}}
Kramer, Rory. 2017. ``Defensible Spaces in Philadelphia: Exploring
Neighborhood Boundaries Through Spatial Analysis.'' \emph{RSF: The
Russell Sage Foundation Journal of the Social Sciences} 3(2):81.

\hyperdef{}{ref-Krysan:2009gm}{\label{ref-Krysan:2009gm}}
Krysan, Maria and Michael D. M. Bader. 2009. ``Racial Blind Spots:
Black-White-Latino Differences in Community Knowledge.'' \emph{Social
Problems} 56(4):677--701.

\hyperdef{}{ref-Lee:2008gm}{\label{ref-Lee:2008gm}}
Lee, Barrett A., Sean F. Reardon, Glenn Firebaugh, Chad R. Farrell,
Stephen A. Matthews, and David O'Sullivan. 2008. ``Beyond the Census
Tract: Patterns and Determinants of Racial Segregation at Multiple
Geographic Scales.'' \emph{American Sociological Review} 73(5):766--91.

\hyperdef{}{ref-Lichter:2015gz}{\label{ref-Lichter:2015gz}}
Lichter, D. T., D. Parisi, and M. C. Taquino. 2015. ``Toward a New
Macro-Segregation? Decomposing Segregation within and between
Metropolitan Cities and Suburbs.'' \emph{American Sociological Review}
80(4):843--73.

\hyperdef{}{ref-Logan:2013in}{\label{ref-Logan:2013in}}
Logan, John R. 2013. ``The Persistence of Segregation in the 21st
Century Metropolis.'' \emph{City \& Community} 12(2):160--68.

\hyperdef{}{ref-Low:2001wn}{\label{ref-Low:2001wn}}
Low, Setha M. 2001. ``The edge and the center: Gated communities and the
discourse of urban fear.'' \emph{American Anthropologist} 103(1):45--58.

\hyperdef{}{ref-Massey:2016cb}{\label{ref-Massey:2016cb}}
Massey, Douglas S. 2016. ``Residential Segregation is the Linchpin of
Racial Stratification.'' \emph{City \& Community} 15(1):4--7.

\hyperdef{}{ref-Massey:1988tq}{\label{ref-Massey:1988tq}}
Massey, Douglas S. and Nancy A. Denton. 1988. ``The Dimensions of
Residential Segregation.'' \emph{Social Forces} 67(2):281--315.

\hyperdef{}{ref-Massey:1993ue}{\label{ref-Massey:1993ue}}
Massey, Douglas S. and Nancy A. Denton. 1993. \emph{American Apartheid:
Segregation and the Making of the Underclass}. Cambridge, Mass. :
Harvard University Press.

\hyperdef{}{ref-Massey:2009fu}{\label{ref-Massey:2009fu}}
Massey, Douglas S., Jonathan Rothwell, and Thurston Domina. 2009. ``The
Changing Bases of Segregation in the United States.'' \emph{The ANNALS
of the American Academy of Political and Social Science} 626(1):74--90.

\hyperdef{}{ref-Matthews:2010eu}{\label{ref-Matthews:2010eu}}
Matthews, Stephen A. 2011. ``Spatial Polygamy and the Heterogeneity of
Place: Studying People and Place via Egocentric Methods.'' Pp. 35--55 in
\emph{Communities, neighborhoods, and health expanding the boundaries of
place}. New York, NY: Springer New York.

\hyperdef{}{ref-Mohl:2002wa}{\label{ref-Mohl:2002wa}}
Mohl, Raymond A. 2002. ``The Interstates and the Cities: Highways,
Housing, and the Freeway Revolt.'' \emph{Poverty and Race Research
Action Council} 1--109.

\hyperdef{}{ref-Morrill:1991vj}{\label{ref-Morrill:1991vj}}
Morrill, Richard L. 1991. ``On the Measure of Geographic Segregation.''
\emph{Geography Research Forum} 11(1):25--36.

\hyperdef{}{ref-R-RColorBrewer}{\label{ref-R-RColorBrewer}}
Neuwirth, Erich. 2014. \emph{RColorBrewer: ColorBrewer Palettes}.

\hyperdef{}{ref-Openshaw:1984wj}{\label{ref-Openshaw:1984wj}}
Openshaw, S. 1984. ``Ecological Fallacies and the Analysis of Areal
Census-Data.'' \emph{Environment and Planning A} 16(1):17--31.

\hyperdef{}{ref-Openshaw:1979wg}{\label{ref-Openshaw:1979wg}}
Openshaw, Stanley and Peter Taylor. 1979. ``A Million or So Correlation
Coefficients: Three Experiments on the Modifiable Area Unit Problem.''
Pp. 127--44 in \emph{Statistical applications in the spatial sciences}.
London : Pion.

\hyperdef{}{ref-OSullivan:2007wo}{\label{ref-OSullivan:2007wo}}
O'Sullivan, David and David W. S. Wong. 2007. ``A Surface-Based Approach
to Measuring Spatial Segregation.'' \emph{Geographical Analysis}
39(2):147--68.

\hyperdef{}{ref-Parisi:2011wk}{\label{ref-Parisi:2011wk}}
Parisi, Domenico, Daniel T. Lichter, and Michael C. Taquino. 2011.
``Multi-Scale Residential Segregation: Black Exceptionalism and
America's Changing Color Line.'' \emph{Social Forces} 89(3):829--52.

\hyperdef{}{ref-R-sp}{\label{ref-R-sp}}
Pebesma, Edzer and Roger S. Bivand. 2015. \emph{sp: Classes and Methods
for Spatial Data}.

\hyperdef{}{ref-Quillian:2012dr}{\label{ref-Quillian:2012dr}}
Quillian, Lincoln. 2012. ``Segregation and Poverty Concentration: The
Role of Three Segregations.'' \emph{American Sociological Review}
77(3):354--79.

\hyperdef{}{ref-R-base}{\label{ref-R-base}}
R Core Team. 2014. \emph{R: A Language and Environment for Statistical
Computing}. Vienna, Austria: R Foundation for Statistical Computing.

\hyperdef{}{ref-Reardon:2011kz}{\label{ref-Reardon:2011kz}}
Reardon, Sean F. and Kendra Bischoff. 2011. ``Income Inequality and
Income Segregation.'' \emph{American Journal of Sociology}
116(4):1092--1153.

\hyperdef{}{ref-Reardon:2002vt}{\label{ref-Reardon:2002vt}}
Reardon, Sean F. and Glenn Firebaugh. 2002. ``Measures of Multigroup
Segregation.'' \emph{Sociological Methodology} 32:33--67.

\hyperdef{}{ref-Reardon:2004vl}{\label{ref-Reardon:2004vl}}
Reardon, Sean F. and David O'Sullivan. 2004. ``Measures of Spatial
Segregation.'' \emph{Sociological Methodology} 34(1):121--62.

\hyperdef{}{ref-Reardon:2009kq}{\label{ref-Reardon:2009kq}}
Reardon, Sean F., Chad R. Farrell, Stephen A. Matthews, David
O'Sullivan, Kendra Bischoff, and Glenn Firebaugh. 2009. ``Race and Space
in the 1990s: Changes in the Geographic Scale of Racial Residential
Segregation, 1990--2000.'' \emph{Social Science Research} 38(1):55--70.

\hyperdef{}{ref-Reardon:2008wa}{\label{ref-Reardon:2008wa}}
Reardon, Sean F., Stephen A. Matthews, David O'Sullivan, Barrett A. Lee,
Glenn Firebaugh, Chad R. Farrell, and Kendra Bischoff. 2008. ``The
Geographic Scale of Metropolitan Racial Segregation.'' \emph{Demography}
45(3):489--514.

\hyperdef{}{ref-R-foreach}{\label{ref-R-foreach}}
Revolution Analytics and Steve Weston. 2014. \emph{foreach: Foreach
looping construct for R}.

\hyperdef{}{ref-Roberto:2016th}{\label{ref-Roberto:2016th}}
Roberto, Elizabeth. 2016. ``The Divergence Index: A Decomposable Measure
of Segregation and Inequality.'' \emph{ArXiv.org}. Retrieved
(http://arxiv.org/abs/1508.01167v2).

\hyperdef{}{ref-Sampson:2008vb}{\label{ref-Sampson:2008vb}}
Sampson, Robert J. and Patrick Sharkey. 2008. ``Neighborhood selection
and the social reproduction of concentrated racial inequality.''
\emph{Demography} 45(1):1--29.

\hyperdef{}{ref-Schindler:2015uv}{\label{ref-Schindler:2015uv}}
Schindler, Sarah. 2015. ``Architectural Exclusion: Discrimination and
Segregation Through Physical Design of the Built Environment.''
\emph{The Yale Law Journal} 124(6):1934--2024.

\hyperdef{}{ref-Schnell:2001wra}{\label{ref-Schnell:2001wra}}
Schnell, I. and B. Yoav. 2001. ``The Sociospatial Isolation of Agents in
Everyday Life Spaces as an Aspect of Segregation.'' \emph{Annals of the
Association of American Geographers} 91(4):622--36.

\hyperdef{}{ref-Spielman:2013jw}{\label{ref-Spielman:2013jw}}
Spielman, Seth E. and John R. Logan. 2013. ``Using High-Resolution
Population Data to Identify Neighborhoods and Establish Their
Boundaries.'' \emph{Annals of the Association of American Geographers}
103(1):67--84.

\hyperdef{}{ref-Sugrue:2005vb}{\label{ref-Sugrue:2005vb}}
Sugrue, Thomas J. 2005. \emph{The Origins of the Urban Crisis: Race and
Inequality in Postwar Detroit}. Princeton: Princeton University Press.

\hyperdef{}{ref-Suttles:1972ue}{\label{ref-Suttles:1972ue}}
Suttles, Gerald D. 1972. \emph{The Social Construction of Communities}.
Chicago, University of Chicago Press.

\hyperdef{}{ref-Taeuber:1965us}{\label{ref-Taeuber:1965us}}
Taeuber, Karl E. and Alma F. Taeuber. 1965. \emph{Negroes in Cities:
Residential Segregation and Neighborhood Change}. Chicago Aldine Pub.
Co.

\hyperdef{}{ref-Tobler:1979ur}{\label{ref-Tobler:1979ur}}
Tobler, Waldo R. 1979. ``Smooth Pycnophylactic Interpolation for
Geographical Regions.'' \emph{Journal of the American Statistical
Association} 74(367):519--30.

\hyperdef{}{ref-CensusSummary:bZv73ozJ}{\label{ref-CensusSummary:bZv73ozJ}}
U.S. Census Bureau. 2011. ``2010 Census Summary File 1---United
States.''

\hyperdef{}{ref-TIGERLineShap:zxE0ic9D}{\label{ref-TIGERLineShap:zxE0ic9D}}
U.S. Census Bureau. 2012. ``2010 TIGER/Line Shapefiles.''

\hyperdef{}{ref-White:1983uo}{\label{ref-White:1983uo}}
White, Michael J. 1983. ``The Measurement of Spatial Segregation.''
\emph{American Journal of Sociology} 88(5):1008--18.

\hyperdef{}{ref-Wong:2010jn}{\label{ref-Wong:2010jn}}
Wong, David W. S. and Shih-Lung Shaw. 2011. ``Measuring Segregation: an
Activity Space Approach.'' \emph{Journal of Geographical Systems}
13(2):127--45.

\hyperdef{}{ref-Wu:2001uk}{\label{ref-Wu:2001uk}}
Wu, X. B. and D. Z. Sui. 2001. ``An Initial Exploration of a
Lacunarity-Based Segregation Measure.'' \emph{Environment and Planning
B: Planning and Design} 28(3):433--46.

\hyperdef{}{ref-Zenk:2011eo}{\label{ref-Zenk:2011eo}}
Zenk, Shannon N., Amy J. Schulz, Stephen A. Matthews, Angela
Odoms-Young, JoEllen Wilbur, Lani Wegrzyn, Kevin Gibbs, Carol
Braunschweig, and Carmen Stokes. 2011. ``Activity space environment and
dietary and physical activity behaviors: A pilot study.'' \emph{Health
\& Place} 17(5):1150--61.

\clearpage

\section{Figures}\label{figures}

\end{document}